\documentclass[%
reprint,
superscriptaddress,
bibnotes,
amsmath,amssymb,
prb,
floatfix,
]{revtex4-2}

\usepackage{graphicx}% Include figure files
\usepackage{dcolumn}% Align table columns on decimal point
\usepackage{bm}% bold math
\usepackage{hyperref}% add hypertext capabilities
\newcommand{\dif}{\text{d}}
\newcommand{\avg}[1]{\left\langle #1 \right\rangle}

\newcommand{\prt}[2]{\frac{\partial #1}{\partial #2}}
\newcommand{\dert}[3]{\frac{\text{d}^#3 #1}{\text{d} #2^#3}}

\renewcommand{\vec}[1]{\boldsymbol{#1}}

\begin{document}

\preprint{}

\title{Explaining Observed Stability of Excitons in Highly Excited CdSe Nanoplatelets}% Force line breaks with \\

\author{F. Garc\'ia Fl\'orez}
\email{f.garciaflorez@uu.nl}
\affiliation{Institute for Theoretical Physics and Center for Extreme Matter and Emergent Phenomena, Utrecht University, Princentonplein 5, 3584 CC Utrecht, The Netherlands}
\author{Aditya Kulkarni}
%\email{}
\affiliation{Optoelectronic Materials Section, Department of Chemical Engineering, Delft University of Technology, Van der Maasweg 9, 2629 HZ, Delft}%
\author{Laurens D. A. Siebbeles}
\email{l.d.a.siebbeles@tudelft.nl}
\affiliation{Optoelectronic Materials Section, Department of Chemical Engineering, Delft University of Technology, Van der Maasweg 9, 2629 HZ, Delft}%
\author{H. T. C. Stoof}%
\email{h.t.c.stoof@uu.nl}
\affiliation{Institute for Theoretical Physics and Center for Extreme Matter and Emergent Phenomena, Utrecht University, Princentonplein 5, 3584 CC Utrecht, The Netherlands}

\date{\today}% It is always \today, today,

\begin{abstract}
    Two-dimensional electron-hole gases in colloidal semiconductors have a wide variety of applications.
    Therefore, a proper physical understanding of these materials is of great importance.
    In this paper we present a detailed theoretical analysis of the recent experimental results by Tomar \emph{et al.} \cite{tomar2019}, that show an unexpected stability of excitons in CdSe nanoplatelets at high photoexcitation densities.
    Including the screening effects by free charges on the exciton properties, our analysis shows that CdSe nanoplatelets behave very differently from bulk CdSe, and in particular do not show a crossover to an electron-hole plasma in the density range studied experimentally, even though there is substantial overlap between the excitons at the highest densities achieved.
    From our results we also conclude that a quantum degenerate exciton gas is realized in the experiments, which opens the prospect of observing superfluidity in CdSe nanoplatelets in the near future.
\end{abstract}

\maketitle

\tableofcontents
\newpage

\section{\label{sec:introduction}Introduction}

Developing the next generation of optoelectronic devices, such as solar cells, photosensors, light-emitting diodes and lasers, involves a wide range of interdisciplinary approaches: new materials synthesis, experimental characterization of the material properties, and theoretical understanding \cite{kovalenko2015,kagan2016,xia2014,bonaccorso2010,chhowalla2013,gan2013,pospischil2013,splendiani2010,xu2014,mak2010}.
In recent years, advancements in this field have been fueled by improvements of the synthesis processes for several materials, such as transition-metal dichalcogenide monolayers and layered perovskites \cite{wang2012,bie2017,manser2016,mahler2010,murray1993,peng2000,yin2005,ithurria2011}.
Thanks to these improvements, further experimental research on inorganic two-dimensional materials has demonstrated their usefulness for engineering optoelectronic and photovoltaic devices due to their strong confinement, direct bandgap, and much more efficient and better industry-integrated chemical manufacturing processes \cite{ye2015,zhao2018,baugher2014,gan2013a,liu2015,pospischil2014,ross2014,schwarz2014,wu2014}.
An essential requirement for such devices is that they function at room temperature so that they can be integrated with traditional silicon-based transistors.
Hence the focus is mostly on room-temperature controlled designs \cite{li2012,salehzadeh2015,shang2017,wu2015,ye2015,zhao2018,yang2017}.
Moreover, the presence of excitons can be either a blessing or a curse for a particular application, which makes the understanding of the excitonic properties of these semiconducting materials of the utmost importance.
Some of these properties have been known for quite some time, with theoretical studies dating back to the eighties \cite{schmitt-rink1985,schmitt-rink1985a,schmitt-rink1986,kumagai1989,kozlov1996,ding1992}, but more recent studies have considerably expanded upon these ideas \cite{asano2014,chernikov2015,steinhoff2017,rustagi2018}.

In this paper, we consider in detail the exciton physics in photoexcited CdSe nanoplatelets.
In particular, our main goal is to explain the observation of Tomar \emph{et al.} that excitons in CdSe nanoplatelets do not break up into electrons and holes and remain present even at high photoexcitation densities \cite{tomar2019}.
We explain this unexpected observation by showing that in a two-dimensional nanoplatelet the screening of the Coulomb potential by free charges is never sufficiently strong to push the exciton bound state into the electron-hole continuum.
This should be contrasted with the three-dimensional bulk material, in which screening alone is enough to make excitons unbind and to let the exciton gas crossover into an electron-hole plasma near the so-called Mott density.
In this regime, the excitons have a substantial overlap with each other and the exciton bound state disappears from the screened Coulomb potential.

\begin{figure}[h]
    \begin{center}
        \includegraphics[width=\linewidth]{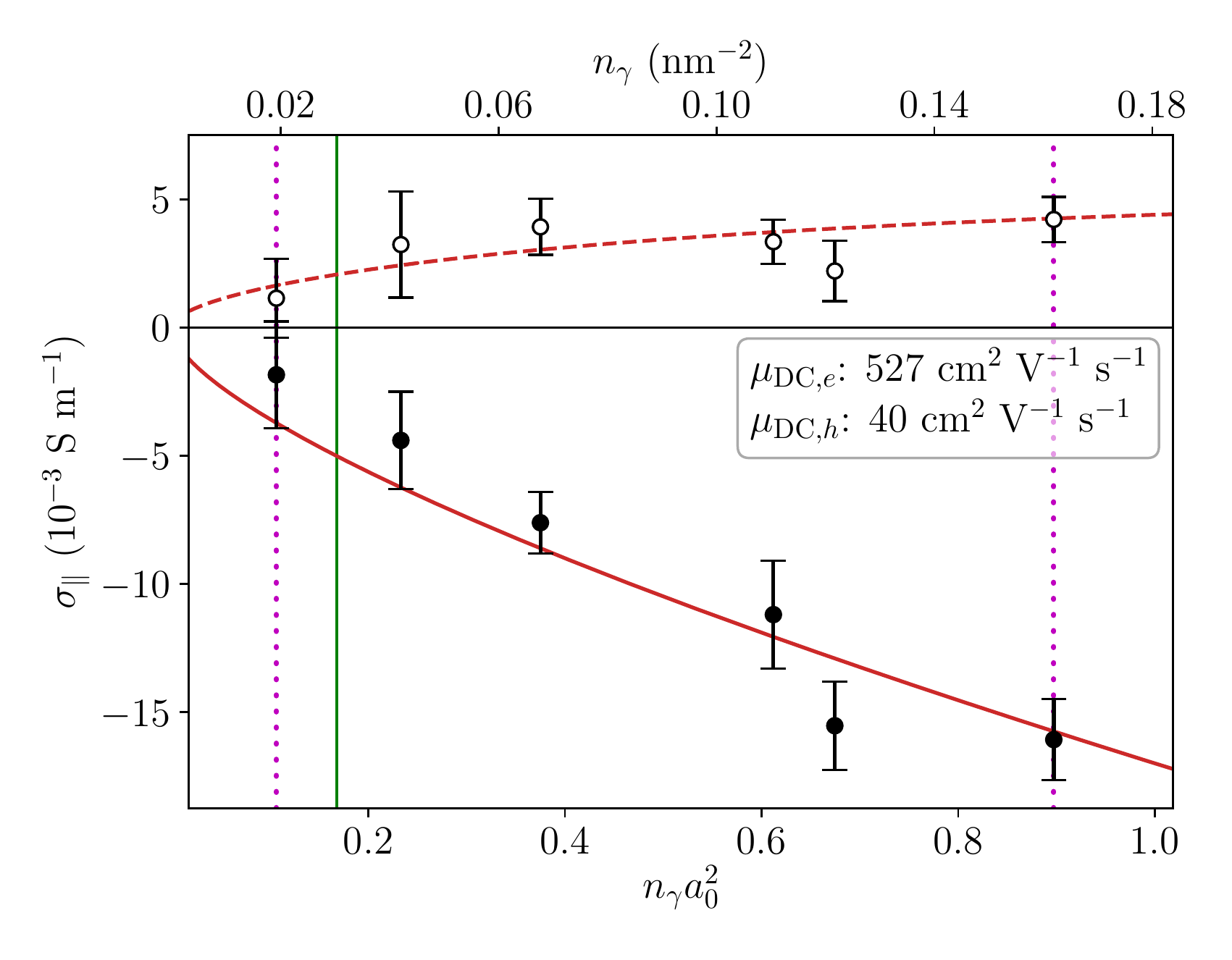}
        \caption{
            The complex in-plane conductivity $\sigma_\parallel$ as a function of photoexcitation density $n_\gamma$, where the dots correspond to the data from Ref. \cite{tomar2019} and the lines to our best model fit.
            Positive values (empty dots and dashed line) give the real part of the conductivity, while the negative values (filled dots and solid line) give the imaginary part.
            Shown in the legend are the numerical results for the electron and hole DC mobilities, which are used to fit the model developed in this paper to the experimental data.
            Dotted magenta vertical lines are shown for reference in other figures.
            The green solid vertical line marks the density above which quantum statistics for excitons needs to be considered.
        }
        \label{fig:cond_fit}
    \end{center}
\end{figure}

To understand the empirical evidence for this better, we show in Fig. \ref{fig:cond_fit} the measured data for both the real and the imaginary parts of the complex in-plane Terahertz (THz) conductivity as a function of the photoexcitation density $n_\gamma$, i.e., the total density of electron-hole pairs initially created by the pump laser and that quickly thermalizes into an (quasi)equilibrium mixture of excitons and free charges.
Since the real part of the THz conductivity is proportional to the density of free charges, and the imaginary part is mainly proportional to the density of excitons, the experimental data clearly shows that at high photoexcitation the density of free charges in the CdSe nanoplatelets saturates, whereas the density of excitons increases linearly with photoexcitation density.
The same conclusion was reached more quantitatively in Ref. \cite{tomar2019} by fitting the measured complex conductivities to the classical equation of state of noninteracting excitons, electrons and holes, also known as the Saha model \cite{saha1920,saha1921}.
Most significantly for our purposes is that the complex THz conductivity does not show the expected crossover from a low-density regime to a high-density regime.
Indeed, if the excitons had become unstable and had unbound into electrons and holes, i.e., if the system had shown a crossover from an exciton gas into an electron-hole plasma near the Mott density, this would have resulted in a relatively sharp change of the complex conductivity.
Clearly, Fig. \ref{fig:cond_fit} does not show such behavior, thus proving that excitons remain stable even at high photoexcitation densities where they start to overlap and $n_\text{exc}a_0^2 \gtrsim 1$, where $n_\text{exc}$ is the density of excitons and $a_0$ is the Bohr radius of the exciton.
As the experimentally reached exciton densities are much higher than the predicted Mott densities in Refs. \cite{rustagi2018,asano2014,steinhoff2017}, a better theoretical understanding is needed, which we aim to provide in this paper.

Our work presents a twofold approach to understand the thermodynamical and charge transport behavior of CdSe nanoplatelets, in which we concentrate on deriving a theoretical model that can be directly compared with the experiments of Ref. \cite{tomar2019}.
However, we anticipate to apply our framework, and extensions thereof, to other two-dimensional materials with similar properties.
As a result, the layoff of the rest of the paper is as follows.
Sec. \ref{sec:experiment} begins by introducing a more detailed description of the system of interest and the measurement process.
Then, Sec. \ref{sec:analysis} presents the various ingredients of our model and analyzes them, by comparing directly with Ref. \cite{tomar2019} and other publications.
Concluding with Sec. \ref{sec:outlook}, we discuss the results obtained in the previous section and show several directions for future endeavors.

\section{\label{sec:experiment}Experimental Setup}

This section introduces the physical characteristics of the CdSe nanoplatelets of interest to us, focusing in Sec. \ref{sec:mat_properties} on the properties of the nanoplatelets and in Sec. \ref{sec:cond_model} on the measurement methodology.
In particular, we show how the complex conductivity of Fig. \ref{fig:cond_fit}, is related to the densities of free charges and excitons in the system.

\subsection{\label{sec:mat_properties}Material Properties}

\begin{figure}[h]
    \begin{center}
        \includegraphics[width=\linewidth]{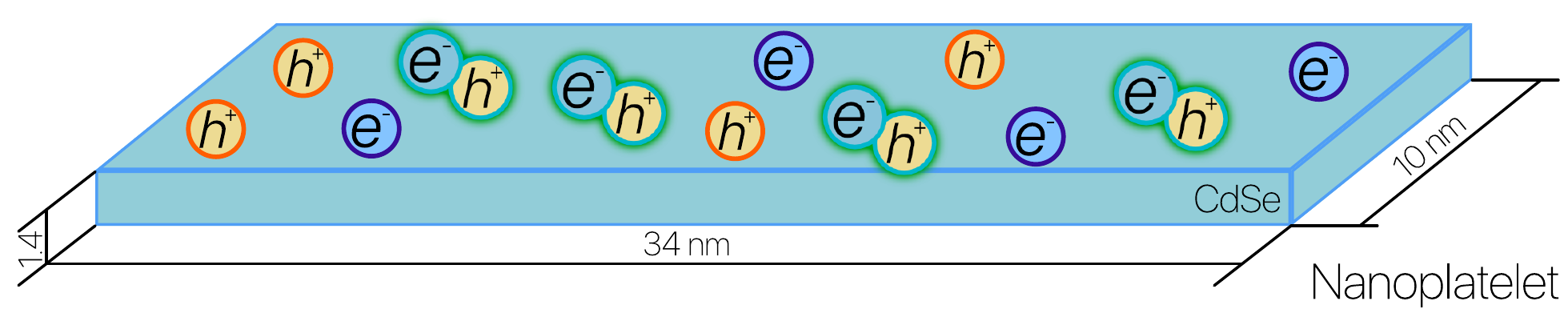}
        \caption{
            Schematic representation of a nanoplatelet, showing their average size in each dimension.
            Confined to the two-dimensional structure are electrons, holes, and excitons that in principle all interact with one another.
        }
        \label{fig:nanoplatelet_schematic}
    \end{center}
\end{figure}

We refer to Sec. 2.1 of Ref. \cite{tomar2019}, and references therein, for a more detailed description of the synthesis procedure of the CdSe nanoplatelets with a thickness of 4.5 monolayers and that in Ref. \cite{benchamekh2014} corresponds to the case of $n=5$ layers.
Our sample is composed of rectangularly-shaped nanoplatelets, shown in Fig. \ref{fig:nanoplatelet_schematic}, with lateral sizes also given in Table \ref{table:parameters}.
The computed band structure of such nanoplatelets has a direct gap at the $\Gamma$-point with the effective masses of electrons and holes, calculated as given in Table \ref{table:parameters} in units of the fundamental electron mass \cite{koster2016}.
Because of the splitting of the hole bands the absorption spectrum exhibits heavy-hole (HH) and light-hole (LH) exciton peaks below the free charges continuum \cite{tomar2019}.
In addition, the finite lateral size leads to the presence of different HH and LH center-of-mass states \cite{richter2017}.
In this work we only consider the lowest HH exciton states as these are mainly populated in the experiments.

Since the nanoplatelets are in solution, the solvent has an effect on the electric field lines outside of the nanoplatelet by substantially decreasing the relative permitivity $\epsilon_r$, compared to its value in bulk CdSe, and thus increasing the strength of the Coulomb interactions between the charges inside the CdSe material.
For simplicity we do not determine the relative permitivity from first principles, instead we compute it from the experimental exciton energy level as described in Sec. \ref{sec:screening}.
The whole system is always kept at room temperature, that is, $T = 294$ K, and most of our discussion will focus on this value.
In subsequent experiments, however, the temperature may be lowered so that new regimes may be explored.

\begin{center}
    \begin{table}[h]
        \begin{tabular}{lccr}
            \hline
            \hline
            \textbf{Name} & \textbf{Symbol} & \textbf{Value} & \textbf{Source} \\
            \hline
            Nanoplatelet $x$-size & $L_x$ & $34\pm1.2$ nm & \cite{tomar2019} \\
            Nanoplatelet $y$-size & $L_y$ & $9.6\pm0.6$ nm & \cite{tomar2019} \\
            Nanoplatelet $z$-size & $L_z$ & $1.37$ nm & \cite{tomar2019} \\
            Nanoplatelet surface & $S_{\text{NPL}}$ & $326$ nm$^2$ & \cite{tomar2019} \\
            Effective electron mass & $m_e$ & $0.22 m_0$ & \cite{benchamekh2014} \\
            Effective hole mass & $m_h$ & $0.41 m_0$ & \cite{benchamekh2014} \\
            Exciton energy level & $E_B^\text{Cou}$ & $-193\pm5$ meV & \cite{tomar2019} \\
            Relative permitivity & $\epsilon_r$ & $6.36$ & Eq. (\ref{eq:cou_eb}) \\
            Temperature & $T$ & $294$ K & \cite{tomar2019} \\
            Peak probe frequency & $\omega_{\text{peak}} / 2\pi$ & $0.9$ THz & \cite{tomar2019} \\
            Exciton Bohr radius & $a_0$ & $2.35$ nm & Eq. (\ref{eq:bohr_def}) \\
            Sat. screening length & $\lambda_s^\text{Sat}$ & $1.68$ nm & Eq. (\ref{eq:scr_length_zero_2d}) \\
            Saturated energy level & $E_B^\text{Sat}$ & $-44.6$ meV & Eq. (\ref{eq:eh_schr_u}) \\
            \hline
            \hline
        \end{tabular}
        \caption{
            Table aggregating the system parameters that are used throughout the paper.
        }
        \label{table:parameters}
    \end{table}
\end{center}

\subsection{\label{sec:cond_model}THz Conductivity of Nanoplatelets}

\begin{figure}[h]
    \begin{center}
        \includegraphics[width=\linewidth]{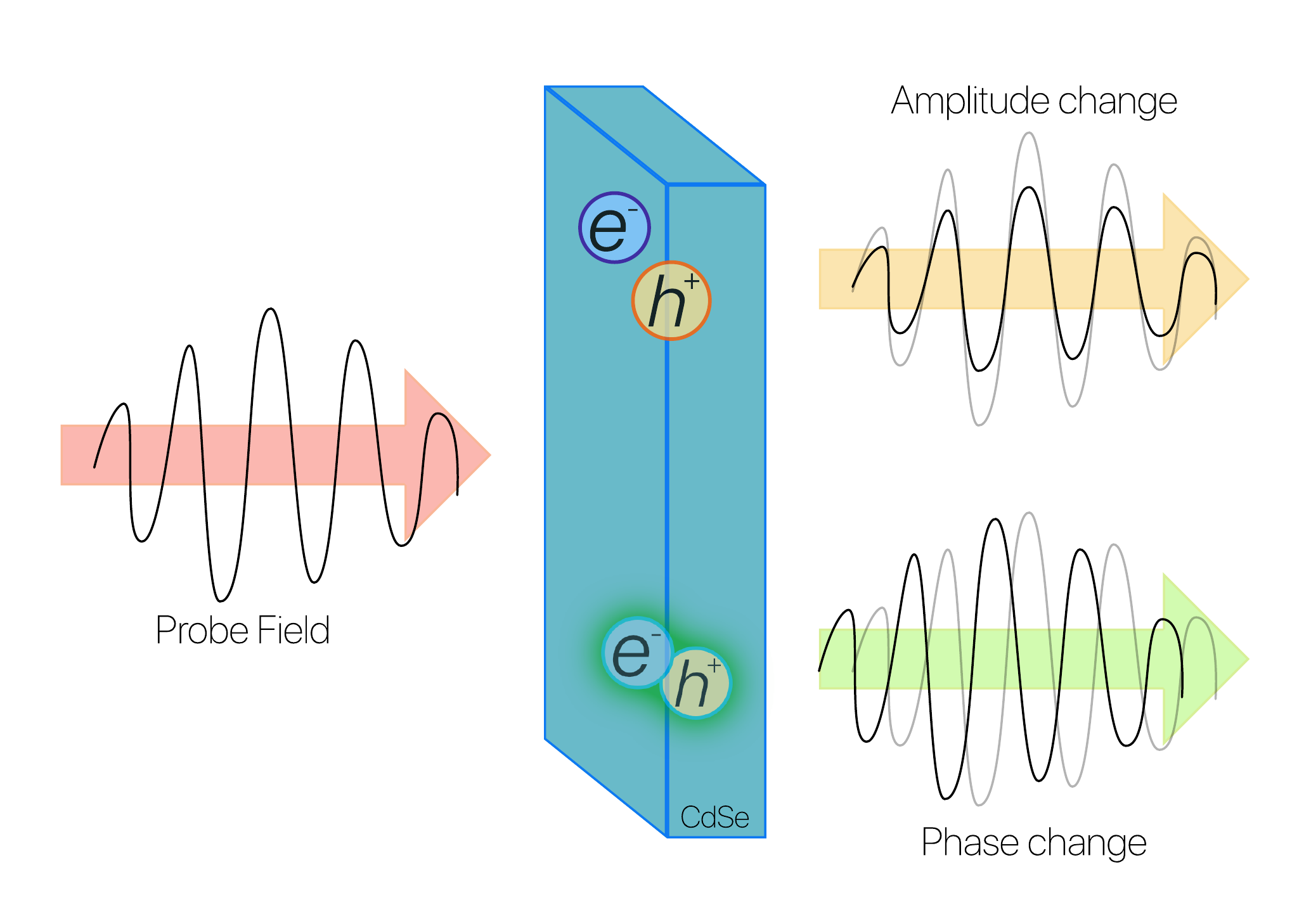}
        \caption{
            When a probe field is applied on a nanoplatelet, free charges and excitons contribute in different ways to the complex conductivity.
            The measured amplitude and phase changes of the THz field due to photoexcitation are determined by the densities of free charges and excitons.
        }
        \label{fig:probe_field}
    \end{center}
\end{figure}

Quantitatively understanding the behavior of free charges and excitons involves comparing our theoretical predictions with measurements.
Experimentally, the sample is optically excited using a pump laser after which a transmitted THz probe pulse is detected to determine the in-plane complex conductivity.
In particular, free charges contribute to the real part of the conductivity $\sigma_{\mathcal{R}}(\omega)$ by absorbing part of the THz probe field, thus reducing its amplitude.
On the other hand, elastic scattering of free charges and the polarizability of the excitons contribute to the imaginary part $\sigma_{\mathcal{I}}(\omega)$ as dephasing of the THz probe field.
Figure \ref{fig:probe_field} shows a schematic representation of these two effects.

Both effects can be computed in linear response and then depend on the real and imaginary parts of the sum of the electron and holes mobilities, $\mu_{\mathcal{R}}(\omega)$ and $\mu_{\mathcal{I}}(\omega)$ respectively, and on the exciton polarizability $\alpha$, i.e., the field-induced dipole moment per unit of field strength.
For convenience, we define the complex mobility as $\mu(\omega) \equiv \mu_{\mathcal{R}}(\omega) + i \mu_{\mathcal{I}}(\omega)$.
Note that the mobilities and polarizability in our case involve only the in-plane response of the nanoplatelets.
As a result, $\sigma_{\mathcal{R}}(\omega)$ and $\sigma_{\mathcal{I}}(\omega)$ are given by \cite{tomar2019,evers2015,cunningham2013,lauth2016}

\begin{align}
    \label{eq:cond_model_real}
    \sigma_{\mathcal{R}}(\omega) &= \frac{2}{3} \frac{e N_\gamma}{V} ~ \mu_{\mathcal{R}}(\omega) ~ \frac{n_q}{n_\gamma} ~~, \\
    \label{eq:cond_model_imag}
    \sigma_{\mathcal{I}}(\omega) &= \frac{2}{3} \frac{e N_\gamma}{V} \left( \mu_{\mathcal{I}}(\omega) ~ \frac{n_q}{n_\gamma} +  \alpha ~ \frac{\omega}{e} \frac{n_{\text{exc}}}{n_\gamma} \right) ~~,
\end{align}

\noindent where $n_q$ is half the density of free charges, i.e., the density of the electrons or of the holes, $n_\text{exc}$ is the density of excitons, $n_\gamma\equiv n_q + n_\text{exc}$ is the density of photoexcitations, and $N_\gamma$ is the total number of photons absorbed in a solution of nanoplatelets with volume $V$.
The ratio $n_q / n_\gamma$ is denoted in Ref. \cite{tomar2019} as the quantum yield $\phi(n_\gamma)$.
The complex conductivity is, similarly as for the mobility, defined as $\sigma(\omega) \equiv \sigma_{\mathcal{R}}(\omega) + i \sigma_{\mathcal{I}}(\omega) \equiv \frac{2}{3}\sigma_\parallel(\omega)$.
Notice that there is a factor of $2/3$ that accounts for the random distribution of nanoplatelet orientations in the solution, meaning that $\mu(\omega)$ and $\alpha$ are the values of the mobility and polarizability for the case in which the THz electric field is parallel to the plane of the nanoplatelets.

\section{\label{sec:analysis}Theoretical Analysis}

Having introduced the details of the system relevant to the experiments, we move on to its theoretical description.
Our goal is to present the simplest possible thermodynamical description that correctly reproduces the conductivity results of Ref. \cite{tomar2019}.
Fundamentally, we are interested in deriving an equation of state involving the three species in the system, i.e., electrons, holes, and their bound exciton state.
Sec. \ref{sec:equilibrium} discusses each of the three components involved, which are put together to solve the equation of state in Sec. \ref{sec:eq_state}.
We then present in Sec. \ref{sec:supp_models} two secondary models to compute the parameters involved in the conductivity calculation of Sec. \ref{sec:cond_model} other than the densities of free charges and excitons, that is, the exciton polarizability and the mobility of free charges.
Finally, using all of the elements introduced, we fit the conductivity measurements, the result of which has already been shown in Fig. \ref{fig:cond_fit}, and compare the fitted DC mobilities of electrons and holes with known values in the literature.

Before starting our discussion, let us consider some aspects that broadly influence this section.
First, the dimensionality of the system.
Even though the platelets used have a nonzero thickness, in our model free charges and excitons behave as if they were contained in a purely two-dimensional plane.
Physically we can interpret this assumption as saying that any excitation in the perpendicular direction has an energy that is too high to affect the dynamics, or equivalently the THz conductivity in that direction.
Second, we focus on the electron-hole dynamics around the $\Gamma$-point of the Brillouin zone, and the dispersion relations are described by the usual parabolic expressions.
Third, free charges can influence each other via a screened Coulomb potential, presented in Sec. \ref{sec:screening}.
Consequently, the Coulomb attraction between electrons and holes can lead to the formation of excitons, whose properties change depending on how strongly the potential is screened.
In principle all charges interact with one another, but because of simplicity only attractive electron-hole interactions are considered.
The effects of electron-electron and hole-hole repulsion have been studied in the literature \cite{ugeda2014}, giving rise to a band-gap renormalization effect.
Since we only consider an equal density of electrons and holes, however, this effect does not occur within the Random Phase Approximation (RPA) on which our model is based.

\subsection{\label{sec:equilibrium}Chemical Equilibrium}

The densities of free charges $n_q$ and excitons $n_\text{exc}$ is determined by the thermal equilibrium between the species, which we can represent as

\begin{equation} \label{eq:system_equilibrium}
    \text{e}^- + \text{h}^+ \rightleftharpoons \text{exc}
    ~~.
\end{equation}

\noindent Expressed in terms of their chemical potentials this implies that $\mu_{\text{e}} + \mu_{\text{h}} = \mu_{\text{exc}}$.
Furthermore, these species are treated thermodynamically as ideal gases, meaning that each density is related to the corresponding chemical potential by

\begin{align}
    \begin{split}
        \label{eq:id_gas_n_q}
        n_q &= g_s \frac{m_e k_B T}{2\pi \hbar^2} \ln\left(1 + e^{\mu_{\text{e}} / k_B T}\right) \\
            &= g_s \frac{m_h k_B T}{2\pi \hbar^2} \ln\left(1 + e^{\mu_{\text{h}} / k_B T}\right) ~~,
    \end{split} \\
    \label{eq:id_gas_n_exc}
    n_{\text{exc}} &= -g_s^2 \frac{(m_e + m_h) k_B T}{2\pi \hbar^2} \ln\left(1 - e^{\left[\mu_{\text{exc}} - E_B(n_q)\right] / k_B T}\right)
    ~~,
\end{align}

\noindent where $g_s = 2$ is the number of spin degrees of freedom of electrons ($\uparrow$ and $\downarrow$) and holes ($\Uparrow$ and $\Downarrow$), $m_e$ and $m_h$ are the effective masses of the electron and hole, $k_B$ is the Boltzmann constant, and $T$ is the temperature.
Note that excitons carry a degeneracy factor of $g_s^2 = 4$ due to the four spin combinations of the constituent electron-hole pair: $\uparrow\Uparrow$, $\uparrow\Downarrow$, $\downarrow\Uparrow$, and $\downarrow\Downarrow$.
We compare our model with measured data by setting $n_\gamma\equiv n_q + n_\text{exc}$ equal to the experimental values, and then finding the densities $n_q$ and $n_\text{exc}$ that satisfy the chemical equilibrium condition.

Notice the energy of the exciton in the exponent of Eq. (\ref{eq:id_gas_n_exc}), which needs to be included because the exciton dispersion relation is

\begin{equation} \label{eq:exc_disp_rel}
    \epsilon_{\vec{K}}^{\text{exc}} = \frac{\hbar^2 \vec{K}^2}{2(m_e + m_h)} - E_B(n_q)
    ~~,
\end{equation}

\noindent whereas the electron and hole dispersions are $\epsilon_{\vec{k}}^e = {\hbar^2 \vec{k}^2}/{2 m_e}$ and $\epsilon_{\vec{k}}^h = {\hbar^2 \vec{k}^2}/{2 m_h}$, respectively.
Moreover, because $n_e = n_h \equiv n_q$, we can derive from Eq. (\ref{eq:id_gas_n_q}) a relation between $\mu_{\text{e}}$ and $\mu_{\text{h}}$ as

\begin{equation} \label{eq:chem_pot_rel}
    \mu_{\text{h}}(\mu_{\text{e}}) = k_B T \ln\left(\left(1 + e^{\mu_{\text{e}} / k_B T}\right)^{\frac{m_e}{m_h}} - 1\right)
    ~~,
\end{equation}

\noindent where we have used

\begin{equation} \label{eq:chem_pot_density}
    \mu_{\alpha}(n_q) = k_B T \ln\left( e^{\frac{\pi\hbar^2}{m_\alpha k_B T}n_q} - 1 \right)
\end{equation}

\noindent for $\alpha = e,h$.

\subsubsection{\label{sec:excitons}Bound Exciton States}

As Sec. \ref{sec:equilibrium} shows, calculating the density of excitons requires knowing their energy $E_B(n_q)$.
Determining this value involves solving the quantum mechanical problem of an electron and a hole mutually attracting each other.
Thus we introduce the exciton wavefunction $\psi_{\text{exc}}(\vec{r})$, where $\vec{r}$ is the relative position and $r\equiv|\vec{r}|$ is its magnitude.
It satisfies the Schr\"odinger equation

\begin{equation} \label{eq:rel_schr_final}
    \left( - \frac{\hbar^2 \vec{\nabla}^2}{2 m_r} + V_\text{sc}(r;n_q) \right) \psi_{\text{exc}}(\vec{r}) = E_B(n_q) \psi_{\text{exc}}(\vec{r})
    ~~,
\end{equation}

\noindent where $m_r = m_e m_h / (m_e + m_h)$ is the reduced mass of the exciton and $V_\text{sc}(r;n_q)$ is the interaction potential.
The energy eigenvalue $E_B(n_q)$ is the negative energy of the exciton, so $-E_B(n_q)$ equals its binding energy.

Due to screening the interaction potential is not a Coulomb potential, but a screened version of it that depends on the density of free charges $n_q$.
Furthermore, it is clear that changes in the potential due to $n_q$ impact the energy of the exciton $E_B(n_q)$.
To solve this equation, we use that the two-dimensional potential in a nanoplatelet is rotationally symmetric to separate the wavefunction in radial and angular parts as $\psi_{\text{exc}}(r,\phi) = u_m(r) r^{-\frac{1}{2}} e^{i m \phi}$.
The angular part is given analytically, while the radial part $u_m(r)$ is a solution of the radial Schr\"odinger equation

\begin{align} \label{eq:eh_schr_u}
    \begin{split}
        &-\frac{\hbar^2}{2 m_r} \dert{}{r}{2} u_m(r) + \left( V_{\text{sc}}(r;n_q) + \frac{\hbar^2}{2 m_r} \frac{m^2 - 1 / 4}{r^2} \right) u_m(r) \\
        &= E_B(n_q) u_m(r) ~~,
    \end{split}
\end{align}

\noindent which is analogous to a one-dimensional Schr\"odinger equation, with an additional contribution to the effective potential next to $V_{\text{sc}}(r;n_q)$ due to the orbital angular momentum of the exciton.
Note that in the following we only treat the $s$-wave case, that is, $m=0$.

Let us first consider the limit $n_q \rightarrow 0$, in which the interaction reduces to the three-dimensional Coulomb potential

\begin{equation} \label{eq:cou_pot}
    \lim_{n_q\rightarrow0} V_{\text{sc}}(r;n_q) = V(r) = -\frac{e^2}{4 \pi \epsilon_0 \epsilon_r} \frac{1}{r}
    ~~,
\end{equation}

\noindent that depends on the relative permitivity $\epsilon_r$ of the solution around the nanoplatelets.
Finding a physical solution to Eq. (\ref{eq:eh_schr_u}) with the Coulomb potential from Eq. (\ref{eq:cou_pot}) fixes the ground-state energy to

\begin{equation} \label{eq:cou_eb}
    \lim_{n_q\rightarrow0} E_B(n_q) = -\frac{2 m_r e^4}{(4 \pi \epsilon_0 \epsilon_r \hbar)^2} \equiv E_B^\text{Cou}
    ~~,
\end{equation}

\noindent which we associate with the lowest bound exciton state.
Note again that $E_B(n_q)$ is related to the exciton binding energy by taking its absolute value $|E_B(n_q)|$.
In our model we only consider an effective relative permitivity $\epsilon_r$ that captures the effect of the solution around the nanoplatelets, that is, the oleate ligands and the hexane solvent.
Instead of determining a value for $\epsilon_r$ based on the physical characteristics of the sample, we simply use the energy of the exciton derived from experiment, given in Table \ref{table:parameters}, and the theoretical expression in Eq. (\ref{eq:cou_eb}) to compute $\epsilon_r$ as

\begin{equation} \label{eq:cou_eb_eps_r}
    \epsilon_r = \frac{e^2}{4\pi\epsilon_0 \hbar} \sqrt{\frac{2m_r}{-E_B^\text{Cou}}}
    ~~.
\end{equation}

\noindent This value of $\epsilon_r$ we then use in Eq. (\ref{eq:cou_pot}) and in all subsequent developments.

\subsubsection{\label{sec:screening}Screening of the Coulomb Potential}

Expanding on the situation explained above, we introduce now the effect of a background of free charges on the electron-hole pair.
Note that, as mentioned previously, the electric field due to charges is not confined to the plane of the nanoplatelet, rather the field lines penetrate the solution around it.
However, since free charges are contained inside the nanoplatelets, screening effects are much less significant compared to those in three-dimensional systems as we show shortly.

Another effect on the potential that may be considered is the finite thickness of the nanoplatelets, which has been extensively studied in the literature \cite{keldysh1979}.
The resulting Keldysh potential turns out to have a logarithmic behavior at short distances and the $1 / r$ Coulomb tail at large distances.
Corrections due to the Keldysh potential are not important for us because the size of the exciton, given by the Bohr radius $a_0$, is typically larger than the height of the nanoplatelets $L_z$.
Table \ref{table:parameters} shows the numerical value of both lengths.

Returning to the screening effects due to the background, we base our derivation on the Random Phase Approximation (RPA) \cite{stoof2009}.
Most of the results in Ref. \cite{stoof2009} can be transferred directly to our situation, only with slight modifications due to the dimensionality, two dimensions as opposed to three dimensions, and having both electrons and holes, as opposed to one type of free carriers, contributing to the screening.
As for the dimensionality of the system, the derivation is general enough that there are no significant changes except for the density of states.
Introducing a new species to the screening is also straightforward, because the effects of the electrons and the holes can just be added together.
Therefore, we can write the screened potential in momentum space as

\begin{align}
    \label{eq:scr_pot_kw}
    \begin{split}
        V_{\text{sc}}^{-1}(k;n_q) &= V^{-1}(k) \\
                                  &- \frac{2 \epsilon_0 \epsilon_r}{e^2} \left(\lambda_{s,e}^{-1}(k; n_q) + \lambda_{s,h}^{-1}(k; n_q)\right)
    \end{split}
    ~~.
\end{align}

\noindent Here we have introduced the two-dimensional Fourier transform of Eq. (\ref{eq:cou_pot}) as $V(k)$.
Defining $k\equiv|\vec{k}|$ and $r\equiv|\vec{r}|$, it is calculated as

\begin{equation}
    V(k) = \int\dif^2\vec{r} ~ V(r) e^{i \vec{k}\cdot\vec{r}} = -\frac{e^2}{2 \epsilon_0 \epsilon_r} \frac{1}{k}
    ~~,
\end{equation}

\noindent where $\vec{k}$ corresponds to the transferred relative momentum between the electron and hole due to the attractive interaction.
In addition, $\lambda_{s,\alpha}^{-1}(k;n_q)$ is the momentum-dependent screening length, given by

\begin{align} \label{eq:pi_kw_general}
    \begin{split}
        & \lambda_{s,\alpha}^{-1}(k; n_q) = \\
        &g_s \int \frac{\dif^2 \vec{k'}}{(2\pi)^2} \frac{N_{FD}\left[\epsilon_{\vec{k}+\vec{k'}}^\alpha - \mu_\alpha(n_q)\right] - N_{FD}\left[\epsilon_{\vec{k'}}^\alpha - \mu_\alpha(n_q)\right]}{\epsilon_{\vec{k}+\vec{k'}}^\alpha - \epsilon_{\vec{k'}}^\alpha} \\
    \end{split}
    ~,
\end{align}

\noindent where $\alpha = e,h$, $N_{FD}(\epsilon)$ is the Fermi-Dirac distribution function $1/(1 + e^{\epsilon / k_B T})$, and $\epsilon_{\vec{k}}^\alpha$ is the kinetic energy of the electron and hole with effective mass $m_\alpha$, i.e., $\epsilon_{\vec{k}}^\alpha = \hbar^2 \vec{k}^2 / 2 m_\alpha$.
Because the free charges are fermions, their screening effects are in principle determined using the Fermi-Dirac distribution.
Fortunately, for the comparison with the experiments we can use the classical limit of Eq. (\ref{eq:pi_kw_general}), that is, using the Maxwell-Boltzmann distribution $N_{MB}(\epsilon) = e^{-\epsilon/k_BT}$.
Approximating $N_{FD}(\epsilon)$ by $N_{MB}(\epsilon)$ is only appropriate when $n_q$ is low enough, which is indeed the case in the experiments.
This leads to the analytical result given in Eq. (\ref{eq:scr_length_mb_2d}).

Independent of the classical limit, Eq. (\ref{eq:pi_kw_general}) can be computed quite accurately by applying the so-called long-wavelength approximation.
If the Fermi-Dirac distribution is used this allows us to study quite accurately the behavior of the system at a very high density of free charges, that is, outside of the regime described by the classical limit.
Physically, this approximation assumes that screening effects affect the interaction potential mostly at large distances and at small momenta.
Hence $\lambda_{s,\alpha}^{-1}(k;n_q)$ can be set to its zero-momentum value $\lambda_{s,\alpha}^{-1}(n_q)$, i.e., the momentum-independent screening length as given by

\begin{equation} \label{eq:scr_length_const_2d}
    \lambda_{s,\alpha}^{-1}(n_q) \equiv \lambda_{s,\alpha}^{-1}(k=0;n_q)
    ~~.
\end{equation}

\noindent Substituting then $\lambda_{s,\alpha}^{-1}(k;n_q) \simeq \lambda_{s,\alpha}^{-1}(n_q)$ into Eq. (\ref{eq:scr_pot_kw}) gives

\begin{equation} \label{eq:scr_pot_lwl_2d}
    V_{\text{sc}}(k; n_q) = -\frac{e^2}{2 \epsilon_0 \epsilon_r} \frac{1}{k + \lambda_{s}^{-1}(n_q)}
    ~~,
\end{equation}

\noindent where $\lambda_{s}^{-1}(n_q) \equiv \lambda_{s,e}^{-1}(n_q) + \lambda_{s,h}^{-1}(n_q)$.
Eq. (\ref{eq:scr_pot_lwl_2d}) can be Fourier transformed back to coordinate space analytically, to obtain an expression for $V_{\text{sc}}(r;n_q)$ given in Eq. (\ref{eq:src_pot_lwl_r_2d}).

\subsubsection{\label{sec:eb_results}Exciton Energy Level}

Let us consider now the accuracy of the long-wavelength approximation.
In the most general case, the potential contains the momentum-dependent screening length $\lambda_{s,\alpha}^{-1}(k;n_q)$ given in Eq. (\ref{eq:pi_kw_general}) using $N_{MB}$ instead of $N_{FD}$ if the classical limit applies.
For the long-wavelength approximation, the potential is computed using the momentum-independent screening length $\lambda_{s,\alpha}^{-1}(n_q)$ given in Eq. (\ref{eq:scr_length_const_2d}).

\begin{figure}[h]
    \begin{center}
        \includegraphics[width=\linewidth]{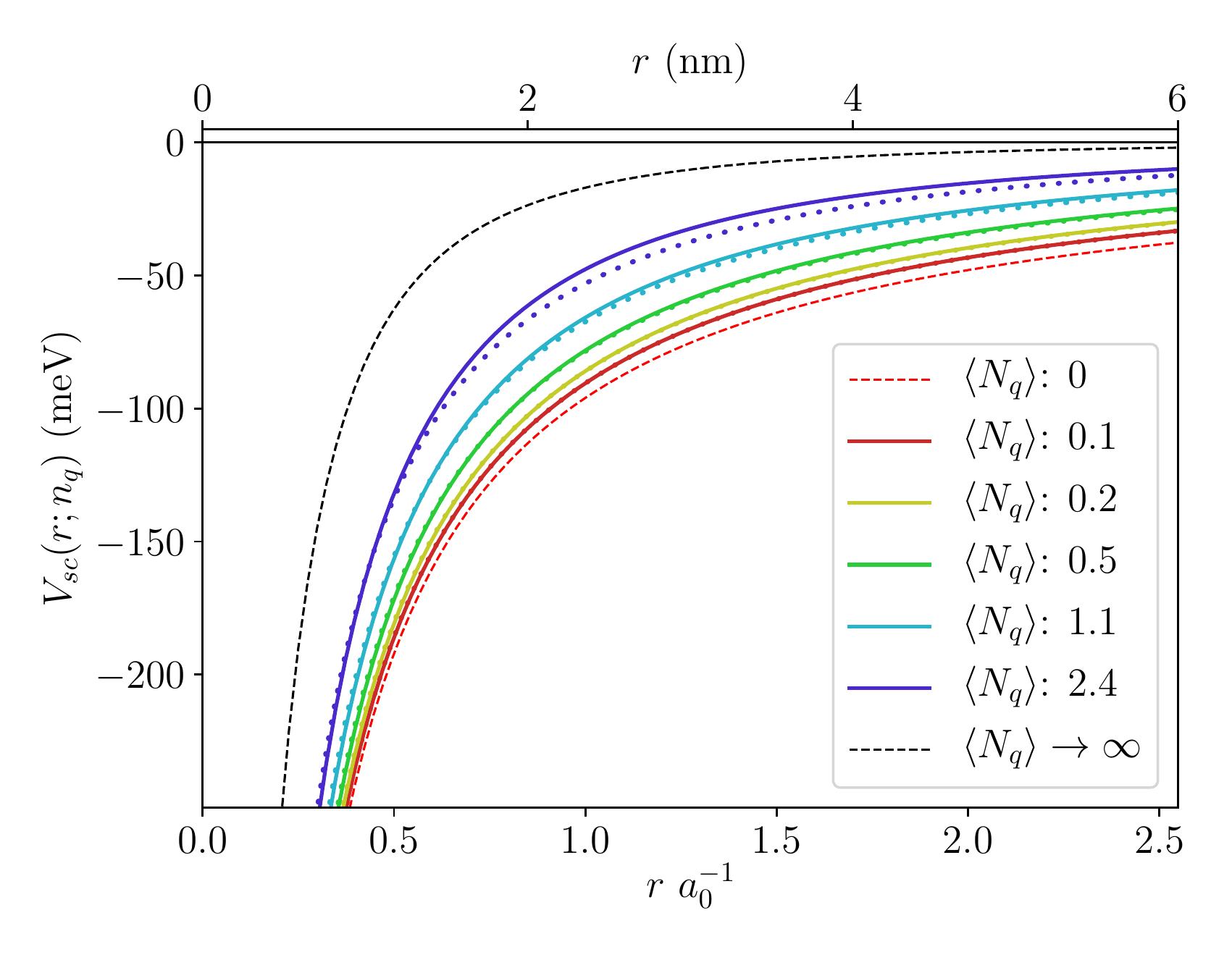}
        \caption{
            Coordinate-space dependence of the screened Coulomb potential for several values of the average number of charges per nanoplatelet $\avg{N_q} \equiv n_q S_{\text{NPL}}$, with $S_{\text{NPL}}$ as given in Table \ref{table:parameters}, and computed for $T=294$ K.
            Lines from bottom to top correspond to low to high value of $\avg{N_q}$.
            Here $a_0$ is the Bohr radius of the exciton, as given in Table \ref{table:parameters}.
            The solid lines are computed using the momentum-dependent screening length, by Fourier transforming Eq. (\ref{eq:scr_pot_kw}), while dotted lines are computed using the screening length for $k\rightarrow0$, by Fourier transforming Eq. (\ref{eq:scr_pot_lwl_2d}), together with Eq. (\ref{eq:scr_length_fd_2d}).
            Included as well are the results for the unscreened Coulomb potential, $\avg{N_q} = 0$ (red dashed line), and the saturated limit, $\avg{N_q}\rightarrow\infty$ (black dashed line).
            As the number of free charges increases, the potential shifts from the Coulomb potential to the saturated potential.
            The finite values of $\avg{N_q}$ used correspond to chemical potentials in the classical limit, that is, satisfying $\mu_{\text{e}}(n_q) / k_B T \lesssim -1$.
        }
        \label{fig:rpot_lwl}
    \end{center}
\end{figure}

\begin{figure}[h]
    \begin{center}
        \includegraphics[width=\linewidth]{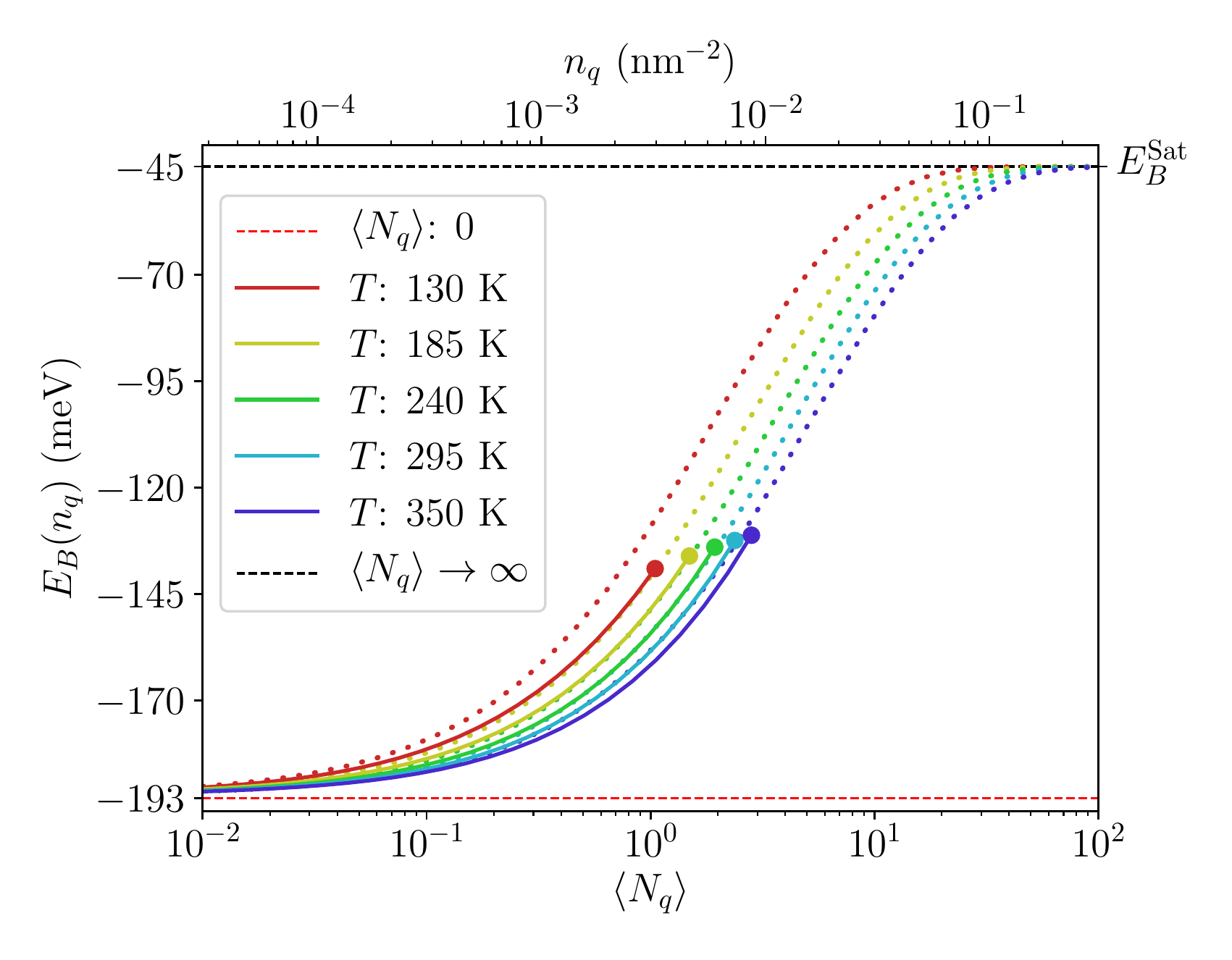}
        \caption{
            Exciton energy as a function of the average number of charges per nanoplatelet $\avg{N_q}\equiv n_q S_{\text{NPL}}$, for different temperatures.
            Lines from left to right correspond to low to high temperatures.
            The value of $S_{\text{NPL}}$ is given in Table \ref{table:parameters}.
            The solid lines are computed using the momentum-dependent screening length, shown in Eq. (\ref{eq:scr_pot_kw}), while dotted lines are computed using the momentum-independent screening length, shown in Eq. (\ref{eq:scr_pot_lwl_2d}), together with Eq. (\ref{eq:scr_length_fd_2d}).
            Solid lines are computed only for the range of $n_q$ that satisfy the classical limit, that is, $\mu_{\text{e}}(n_q) / k_B T \lesssim -1$.
            The point at which the classical limit can no longer considered to be valid, $\mu_{\text{e}}(n_q) / k_B T = -1$, is marked with a dot.
            Dotted lines, corresponding to the long-wavelength approximation, are computed for higher values of $\avg{N_q}$, and the limit $\avg{N_q}\rightarrow\infty$ saturates to $E_B^\text{Sat}$.
            Included as well are the results for the unscreened Coulomb potential, $\avg{N_q} = 0$ (red dashed line), and the saturated limit, $\avg{N_q}\rightarrow\infty$ (black dashed line).
        }
        \label{fig:exc_be}
    \end{center}
\end{figure}

First, Fig. \ref{fig:rpot_lwl} shows the coordinate-space dependence of both potentials, obtained by Fourier transforming Eq. (\ref{eq:scr_pot_kw}).
In the case of the classical limit, the momentum-dependent screening length $\lambda_{s,\alpha}^{-1}(k;n_q)$ is given in Eq. (\ref{eq:scr_length_mb_2d}), while in the long-wavelength approximation it is set to its zero-momentum value $\lambda_{s,\alpha}^{-1}(n_q)$.
As already mentioned, only in the latter case the Fourier transform to coordinate space can be performed analytically, resulting in Eq. (\ref{eq:src_pot_lwl_r_2d}).
In the former case it is computed numerically.
In both Fig. \ref{fig:rpot_lwl} and Fig. \ref{fig:exc_be}, solid lines and dotted lines correspond to the classical limit and long-wavelength approximation, respectively.
Also, the unscreened Coulomb limit ($n_q \rightarrow 0$, red thin dashed line) and the saturated limit ($n_q \rightarrow \infty$, black thin dashed line) are included for reference.
Figure \ref{fig:rpot_lwl} shows the coordinate-space dependence of the screened potential at room temperature, as given in Table \ref{table:parameters}, for different values of the average number of free charges per nanoplatelet $\avg{N_q} \equiv n_q S_\text{NPL}$.
The value of $S_\text{NPL}$ is given in Table \ref{table:parameters} as well.
Even though both potentials seem to be very close, notice that at short distances the long-wavelength approximation underestimates compared to the full classical result, while at long distances it overestimates.
This behavior changes around a distance of $r \simeq a_0 / 4$ and a corresponding energy of $-150$ meV.
As a consequence, the exciton energy level depends somewhat on the potential used, since $E_B^\text{Cou} \simeq -200$ meV, as given in Table \ref{table:parameters}.

Second, using these two potentials to solve the Schr\"odinger equation given in Eq. (\ref{eq:eh_schr_u}) results in two different exciton binding energies.
Figure \ref{fig:exc_be} shows the dependence of the exciton energy as a function of $\avg{N_q}$, for several temperatures.
As expected, for both cases and every temperature, we recover $E_B^\text{Cou}$ for $\avg{N_q} \rightarrow 0$.
However, as $\avg{N_q}$ increases, differences of up to $5$ meV rapidly become apparent.
Since the momentum-independent case underestimates the potential at short distances, the resulting value of $E_B(n_q)$ is less negative than the corresponding one computed with the momentum-dependent screening length.
Notice that in the limit $\avg{N_q}\rightarrow\infty$ the exciton energy saturates to the constant value $E_B^\text{Sat}$.
The value of $E_B^\text{Sat}$ is given in Table \ref{table:parameters}.
Compared to the thermal energy, we find that $\left|E_B^\text{Sat}\right| / k_B T \simeq 3 / 2$, which means that the bound exciton state does not break up due to thermal fluctuations.

Summarizing, it is clear from these room-temperature results that there is a well-defined bound exciton state even at high densities.
Since the exciton also does not break up due to thermal fluctuations, the system does not show a Mott crossover to an electron-hole plasma regime solely due to screening.
Hence, thermal effects are not significant enough at room temperature to affect the exciton state, and only at temperatures around $500$ K the ratio $\left|E_B^\text{Sat}\right| / k_B T$ becomes equal to one.

\subsubsection{\label{sec:sat_scr}Saturated Energy Level}

Let us now focus on understanding better the results that Figs. \ref{fig:rpot_lwl} and \ref{fig:exc_be} show.
These are driven by changes in the screening length, therefore we need to study the dependence of $\lambda_s^{-1}(n_q)$ on the density of free charges $n_q$.
Since the classical limit does not cover the full range of densities, that is, from $n_q \rightarrow 0$ to $n_q \rightarrow \infty$, we instead use only the long-wavelength approximation, that nicely illustrates the essential physics.
In this case, Eq. (\ref{eq:scr_length_const_2d}) can in general be rewritten as

\begin{equation} \label{eq:scr_length_2d}
    \lambda_{s,\alpha}^{-1}(n_q) = \frac{e^2}{2\epsilon_0\epsilon_r} \left.\prt{n_\alpha(\mu)}{\mu}\right|_{\mu=\mu_\alpha(n_q)}
        ~~.
\end{equation}

\noindent The derivation of this result is briefly shown in Sec. 8.7.3 of Ref. \cite{stoof2009}.
Here $n_\alpha(\mu)$, with $\alpha = e,h$, is the density of free charges as a function of the chemical potential $\mu$, as given in Eq. (\ref{eq:id_gas_n_q}).
The chemical potential $\mu = \mu_\alpha(n_q)$ is computed using Eq. (\ref{eq:chem_pot_density}).
Using this result we can discuss in more detail how the inverse screening length $\lambda_s^{-1}(n_q) \equiv \lambda_{s,e}^{-1}(n_q) + \lambda_{s,h}^{-1}(n_q)$ behaves as a function of $n_q$.
For this purpose, we substitute Eq. (\ref{eq:id_gas_n_q}) into Eq. (\ref{eq:scr_length_2d}) to find analytically

\begin{equation} \label{eq:scr_length_fd_2d}
    \lambda_{s,\alpha}^{-1}(n_q) = \frac{e^2}{2\epsilon_0\epsilon_r} \frac{m_\alpha}{\pi\hbar^2} \left( 1 - e^{-\frac{\pi \hbar^2}{m_\alpha k_B T} n_q} \right)
    ~~.
\end{equation}

\begin{figure}[h]
    \begin{center}
        \includegraphics[width=\linewidth]{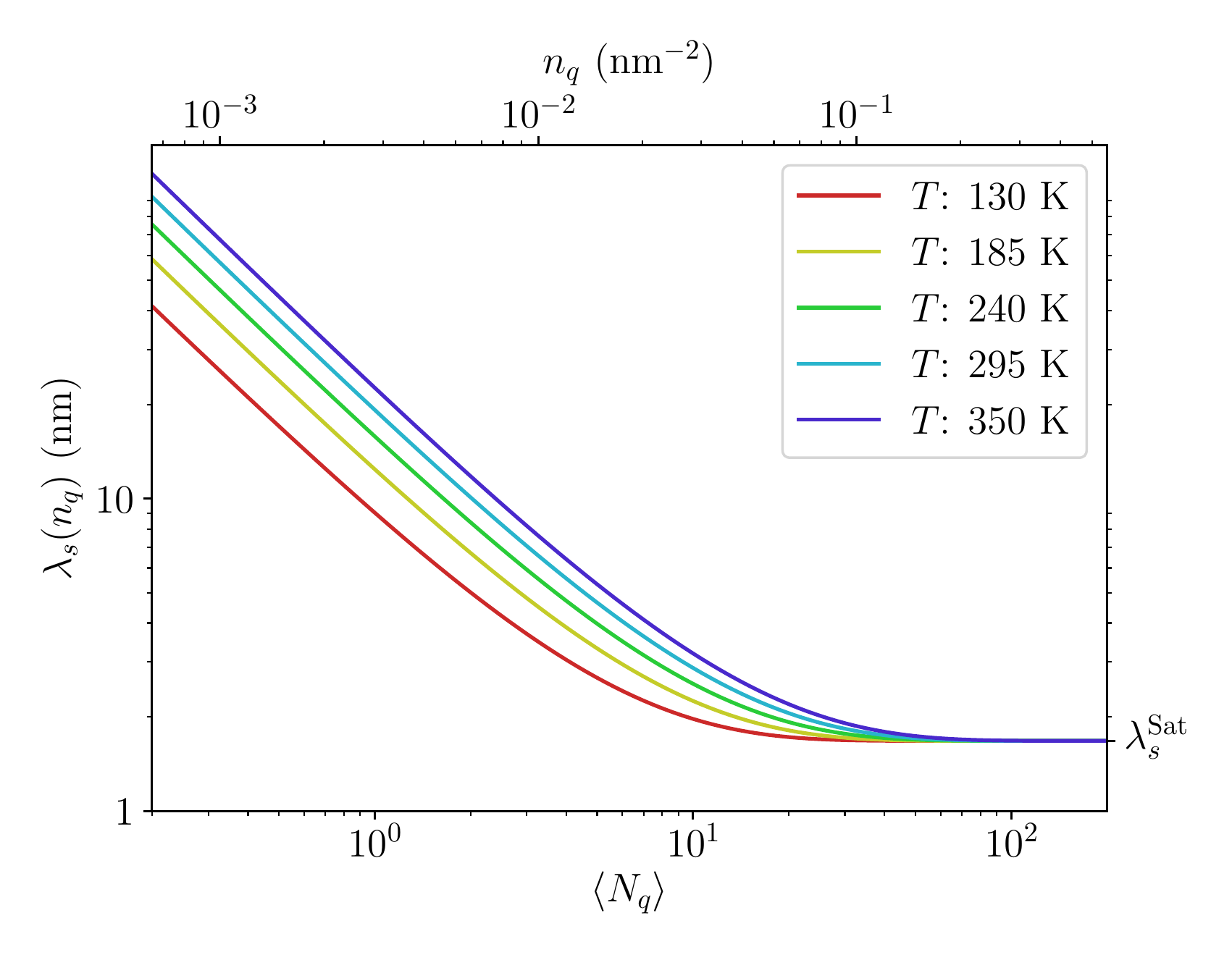}
        \caption{
            Screening length as a function of the average number of free charges per nanoplatelet $\avg{N_q}\equiv n_q S_\text{NPL}$, for different temperatures.
            Lines from left to right correspond to low to high temperatures.
            The screening length $\lambda_s(n_q)$ is computed using Eq. (\ref{eq:scr_length_fd_2d}).
            The value of $S_\text{NPL}$ is given in Table \ref{table:parameters}.
            Notice how in the limit $\avg{N_q}\rightarrow\infty$, $\lambda_s\rightarrow\lambda_s^\text{Sat}$, as Eq. (\ref{eq:scr_length_fd_2d}) describes.
            The value of $\lambda_s^\text{Sat}$ is given in Table \ref{table:parameters}.
        }
        \label{fig:sat_src_length}
    \end{center}
\end{figure}

Figure \ref{fig:sat_src_length} shows the behavior of $\lambda_s(n_q)$ as a function of the average number of free charges per nanoplatelet $\avg{N_q}\equiv n_q S_\text{NPL}$.
The value of $S_\text{NPL}$ is given in Table \ref{table:parameters}.
It is clear that $\lambda_s(n_q)$ always tends toward the same value at high density regardless of the temperature.
The saturation can be immediately seen from Eq. (\ref{eq:scr_length_fd_2d}), since it describes a quantity that converges to a constant value in the limit $n_q \rightarrow \infty$.
Thus, we define the saturated screening length $\lambda_s^\text{Sat}$ as

\begin{equation} \label{eq:scr_length_zero_2d}
    \lim_{n_q\rightarrow\infty} \lambda_s(n_q) = \frac{2\epsilon_0\epsilon_r}{e^2} \frac{\pi \hbar^2}{m_e + m_h} \equiv \lambda_s^{\text{Sat}}
    ~~,
\end{equation}

\noindent that can be written in terms of the unscreened exciton energy as

\begin{equation} \label{eq:scr_length_zero_2d_eb}
    \lambda_s^{\text{Sat}} = \sqrt{\frac{m_r}{-2E_B^\text{Cou}}} \frac{\hbar}{m_e + m_h}
    ~~.
\end{equation}

\noindent This expression does not depend on $n_q$, but only on parameters of the system: the relative permitivity $\epsilon_r$ (or exciton energy $E_B$) and the effective masses of electrons and holes, $m_e$ and $m_h$, respectively.
The main consequence of this result is that, in two dimensions, the screening length does not go to zero as the system becomes dominated by free charges, but instead it tends to a constant.
This is an unexpected result, since it differs greatly from the behavior of three-dimensional materials \cite{deleeuw2016}.
Consider for a moment the more general context of $d$ dimensions and high densities.
In this case, we have that the density of free charges is given by the zero-temperature result $n_\alpha \propto \mu_\alpha^{d/2}$, since it is equal to the volume of the $d$-dimensional Fermi sphere.
Substituting this behavior into Eq. (\ref{eq:scr_length_2d}) we find that $\lambda_{s,\alpha}(n_q)\propto n_q^{2/d-1}$, which goes to zero as $n_q^{-1/3}$ for $d=3$ but to a constant for $d=2$, in agreement with Eqs. (\ref{eq:scr_length_zero_2d}) and (\ref{eq:scr_length_zero_2d_eb}).

Because the screened two-dimensional Coulomb potential thus saturates at a high density of free charges, solving the Schr\"odinger equation given in Eq. (\ref{eq:eh_schr_u}) also shows saturation of the exciton energy.
For convenience, we define the saturated exciton energy $E_B^\text{Sat}$ as

\begin{equation}
    \lim_{n_q\rightarrow\infty}E_B(n_q) \equiv E_B^\text{Sat}
    ~~.
\end{equation}

\noindent Most importantly, if $\left|E_B^\text{Sat}\right|$ is large compared to $k_B T$ then the bound exciton state cannot be broken up by thermal effects.
Consequently, as long as there are no significant other contributions to the exciton energy, two-dimensional systems do not show a crossover to an electron-hole plasma regime due to screening effects alone.

\subsection{\label{sec:eq_state}Equation of State}

\begin{figure}[h]
    \begin{center}
        \includegraphics[width=\linewidth]{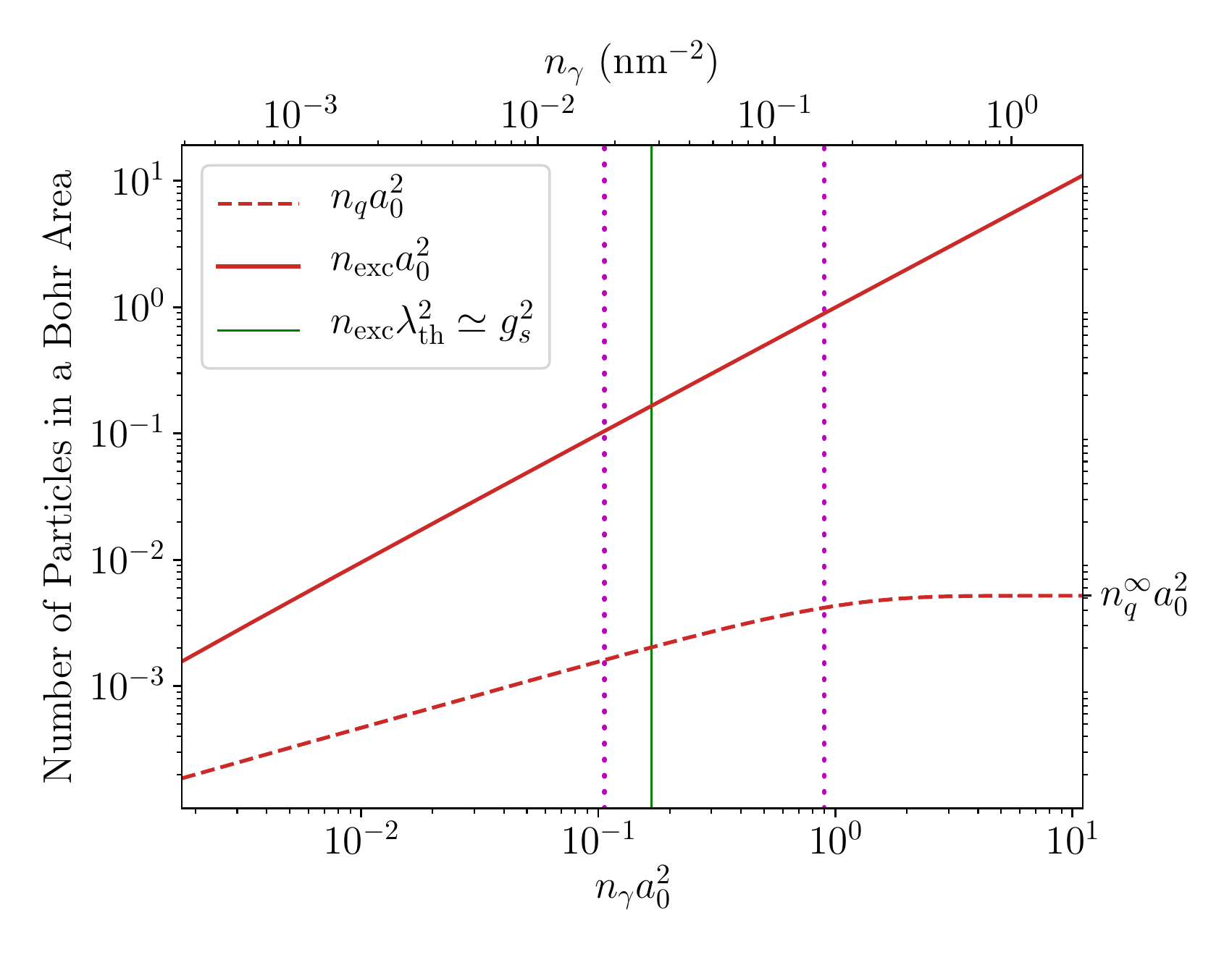}
        \caption{
            Number of free charges $n_q a_0^2$ and excitons $n_{\text{exc}} a_0^2$ in a Bohr area as a function of the number of photoexcitations in a Bohr area $n_\gamma a_0^2$, computed for $T=294$ K.
            The vertical dotted magenta lines mark the range in which we have experimental data.
            Notice that the number of electrons saturates, and at high densities there are mainly only excitons in the system.
            The vertical solid green line represents the density at which quantum effects become significant, further explained in Sec. \ref{sec:qstat_exc}.
            Notice that measurements are taken at higher densities, and thus excitons cannot be considered as classical particles.
        }
        \label{fig:density_result}
    \end{center}
\end{figure}

Now that we have obtained the dependence of the exciton energy level on the density of free charges, we can return to the equation of state and focus on the dependence of the system on the photoexcitation density $n_\gamma$.
Figure \ref{fig:density_result} shows the result of solving Eqs. (\ref{eq:id_gas_n_q}) and (\ref{eq:id_gas_n_exc}) using the parameters given in Table \ref{table:parameters}.
Here, instead of scaling densities with the nanoplatelet surface $S_{\text{NPL}}$, we have used the Bohr area $a_0^2$.
Physically, $n_{\text{exc}}a_0^2 \simeq 1$ marks the limit at which excitons start to overlap, and therefore interactions ought to become more significant.
Notice that the density of excitons is much larger than that of free charges, making the former the dominant species.
Furthermore, as $n_\gamma \rightarrow \infty$, the density of free charges clearly tends toward the saturation value $n_q^\infty$.
As a consequence, it is clear that our model does not predict a Mott crossover for the range of densities explored in experiments.
At even higher densities, we expect exciton-exciton interactions to introduce significant corrections that render our model no longer valid.
Moreover, these interactions could ultimately produce a crossover to an electron-hole plasma regime, although not via the same mechanism based on screening alone as known from three-dimensional systems.
In Sec. \ref{sec:outlook} we revisit this important topic.

\subsubsection{\label{sec:charge_sat}Saturation of Charges}

Let us study in more detail the limit of very high photoexcitation density, that is, $n_\gamma \rightarrow \infty$.
Because our model treats excitons as an ideal gas, $n_\text{exc}$ becomes arbitrarily large as $\mu_{\text{exc}} \rightarrow E_B(n_q)$.
Notice that in two dimensions the chemical potential cannot become equal to the exciton energy, as opposed to three-dimensional bosons that Bose-condense when $\mu_{\text{exc}} = E_B(n_q)$.
Since in this limit $n_\text{exc} \rightarrow \infty$, while $n_q$ is still a finite quantity given by solving

\begin{equation} \label{eq:n_q_saturation}
    n_q^\infty = g_s \frac{m_e k_B T}{2\pi \hbar^2} \ln\left(1 + e^{\left[ E_B(n_q^\infty) - \mu_{\text{h}}(n_q^{\infty}) \right] / k_B T}\right)
    ~~,
\end{equation}

\noindent it is clear that in the high-density regime the physics is dominated by excitons.
Here we have used Eq. (\ref{eq:id_gas_n_q}) and

\begin{equation} \label{eq:mu_e_lim}
    E_B(n_q^{\infty}) - \mu_{\text{e}}(n_q^{\infty}) - \mu_{\text{h}}(n_q^{\infty}) = 0
    ~~.
\end{equation}

Having determined the maximum density of free charges in the system, we can revisit the condition for using the classical limit.
Recall that in order for the classical limit to be valid, the density of free charges has to be low enough.
Specifically, Fig. \ref{fig:density_result} shows that

\begin{equation}
    n_q^\infty \simeq 9.37 \cdot 10^{-4} ~~\text{nm}^{-2}
    ~~,
\end{equation}

\noindent which corresponds to the chemical potentials

\begin{align}
    \label{eq:mu_e_nq_inf}
    \frac{\mu_{\text{e}}(n_q^\infty)}{k_B T} &\simeq -3.19 ~~, \\
    \label{eq:mu_h_nq_inf}
    \frac{\mu_{\text{h}}(n_q^\infty)}{k_B T} &\simeq -3.82 ~~.
\end{align}

\noindent These values are negative enough that we can safely assume in the description of the experiments that the classical limit is valid for the free charges.

\subsubsection{\label{sec:qstat_exc}Quantum Statistics of Excitons}

Since the density of excitons is very high, describing them involves quantum statistics, that is, using the Bose-Einstein distribution instead of the Maxwell-Boltzmann one.
As a reminder, a good estimate of the density above which the Maxwell-Boltzmann approximation is no longer valid is given by

\begin{equation} \label{eq:exc_deg_lim}
    n_{\text{exc}} \lambda_{\text{th}}^2 \simeq g_s^2
    ~~,
\end{equation}

\noindent where $\lambda_{\text{th}}\equiv\hbar\sqrt{2\pi/m_r k_B T}$ is the thermal de Broglie wavelength.
This density is marked in Figs. \ref{fig:cond_fit} and \ref{fig:density_result} with a vertical solid green line.
For higher densities, excitons are actually described by Eq. (\ref{eq:id_gas_n_exc}), and cannot be approximated by assuming that $(\mu_{\text{exc}} - E_B(n_q)) / k_B T \ll 0$.
In other words, they can no longer be considered as classical particles, and thus quantum effects become significant.

On a related note, interacting bosons show a transition to a superfluid phase when their density is high and the temperature is low enough.
However, our model does not include these interactions and this regime is never reached.
Instead, in the ideal case the density of excitons increases arbitrarily high as $\mu_{\text{exc}} \rightarrow E_B(n_q)$.

\subsection{\label{sec:supp_models}Supporting Models}

To be able to confront our equation of state with the THz experiment, we introduce in Secs. \ref{sec:pol_model} and \ref{sec:diff_model} two secondary models for calculating the exciton polarizability and the changes in the mobility of free charges due to finite-size effects, respectively.
In the end we can give a complete picture of the conductivity shown in Fig. \ref{fig:cond_fit} in the introduction.
With the help of these two supporting models we are left with two theoretically unknown parameters, $\mu_{\text{DC},e}$ and $\mu_{\text{DC},h}$, i.e., the DC mobilities of electrons and holes.
These parameters are fitted to reproduce the THz conductivity data and then compared to values in literature for further checks on our results.

\subsubsection{\label{sec:pol_model}Polarizability}

One variable that can be determined \emph{ab initio} is the exciton polarizability.
Using standard results for the polarizability of hydrogen atoms confined to a two-dimensional plane \cite{yang1991,pedersen2007}, we can obtain analytical results valid for our situation.
Note that the attractive Coulomb potential that binds the atom is three-dimensional.
The results of Refs. \cite{yang1991,pedersen2007} can be applied to our situation by introducing a relative permitivity $\epsilon_r$, and setting the reduced mass $m_r$ to that of the exciton.
This leads to a polarizability $\alpha$ given by

\begin{equation} \label{eq:pol_def}
    \alpha = \frac{21}{2^7} ~ 4 \pi \epsilon_0 \epsilon_r ~ a_0^3
    ~~,
\end{equation}

\noindent where $a_0$ is the three-dimensional Bohr radius of the exciton that is defined as

\begin{equation} \label{eq:bohr_def}
    a_0 \equiv 4 \pi \epsilon_0 \epsilon_r ~ \frac{\hbar^2}{m_r e^2}
    ~~.
\end{equation}

\noindent Combining Eqs. (\ref{eq:cou_eb_eps_r}) and (\ref{eq:bohr_def}) results in an expression for the Bohr radius as a function of the exciton energy, i.e., $a_0(E_B)= \sqrt{{2 \hbar^2}/{-E_B m_r}}$.
Further substituting $a_0(E_B)$ into Eq. (\ref{eq:pol_def}) relates the polarizability to the exciton energy $E_B$, as

\begin{align}
    \label{eq:pol_eb}
    \alpha(E_B) &= \frac{21}{2^7} ~ 4 \pi \epsilon_0 \epsilon_r ~ \left( \frac{2 \hbar^2}{-E_B m_r} \right)^{\frac{3}{2}} ~~.
\end{align}

\begin{figure}[h]
    \begin{center}
        \includegraphics[width=\linewidth]{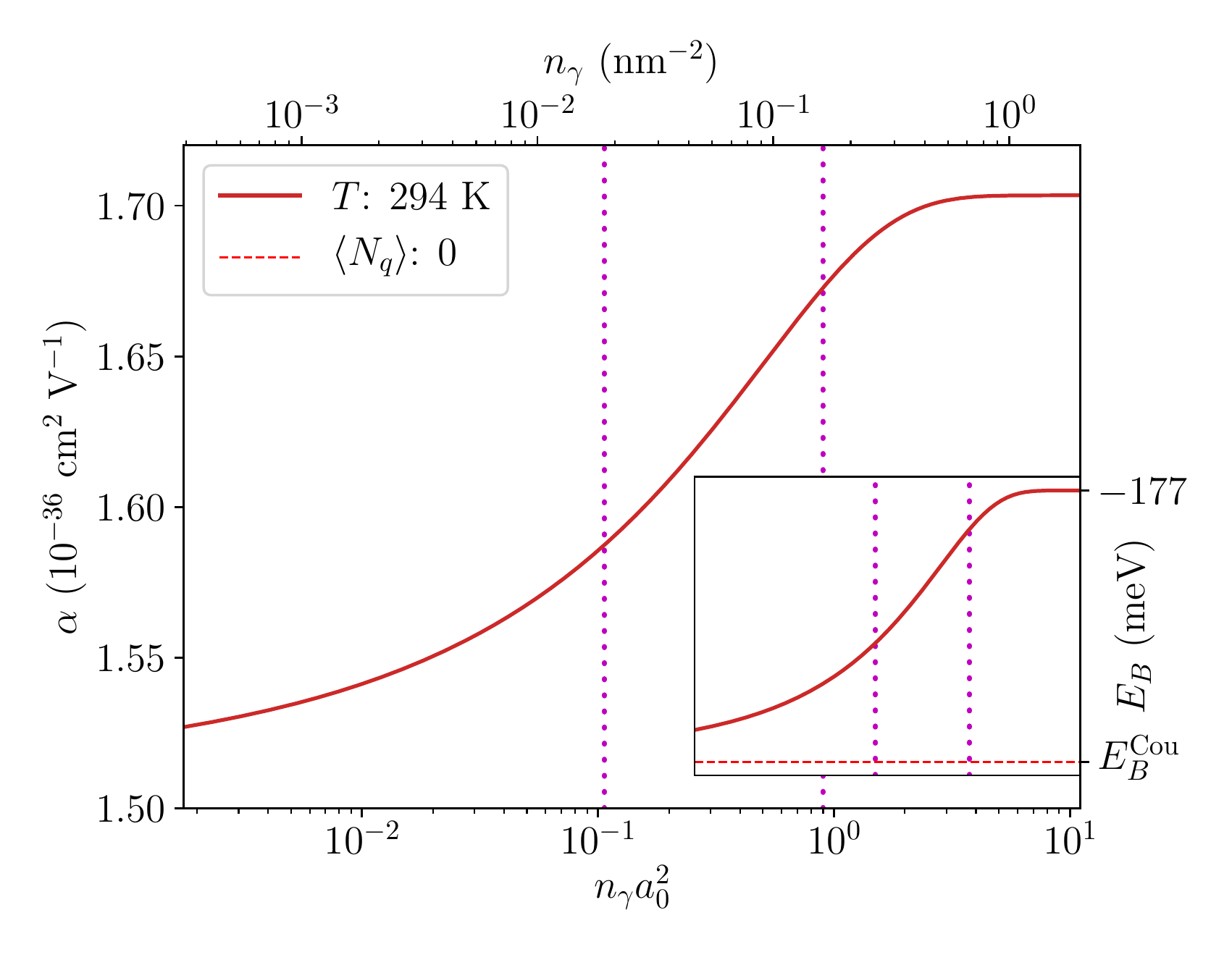}
        \caption{
            Exciton polarizability (main plot) and exciton energy level (inset plot) as a function of the number of photoexcitations in a Bohr area $n_\gamma a_0^2$.
            As explained before, in the limit $n_\gamma\rightarrow\infty$, the density of free charges saturates and consequently $E_B \rightarrow E_B^\infty$.
            The vertical dotted magenta lines mark the range of densities measured in the experiments.
        }
        \label{fig:eb_photo_result}
    \end{center}
\end{figure}

As presented in Sec. \ref{sec:excitons}, the exciton energy level depends on the density of free charges $n_q$.
Thus, using Eq. (\ref{eq:pol_eb}) we can compute the polarizability even when the Coulomb potential is screened.
Implicitly, this procedure assumes that the exciton wavefunction that solves Eq. (\ref{eq:eh_schr_u}), using the Coulomb potential as given in Eq. (\ref{eq:cou_pot}), does not significantly differ compared to solving it using a screened potential $V_\text{sc}(r;n_q)$.
After solving the equation of state, we can thus compute the exciton energy and polarizability for each value of $n_\gamma$.
Figure \ref{fig:eb_photo_result} shows $\alpha(n_\gamma)$ in the main plot, and $E_B(n_\gamma)$ in the inset plot.
Notice that the variation in $E_B(n_\gamma)$ is not very large, $|E_B^\infty - E_B^\text{Cou}| \simeq 16$ meV, where $E_B^\infty \equiv E_B(n_\gamma\rightarrow\infty)$, meaning that the screening is indeed only moderate due to the saturation of the density of free charges.
When $n_\gamma$ is small $E_B(n_\gamma)$ behaves similarly to the results shown in Fig. \ref{fig:exc_be}.
As the photoexcitation density increases the screening effects become more significant, until $n_q$ saturates to $n_q^\infty$ and consequently $E_B(n_\gamma)$ saturates to $E_B^\infty$.
As mentioned, the polarizability as a function of density $\alpha(n_\gamma)$ is computed using Eq. (\ref{eq:pol_eb}), and setting $E_B = E_B(n_\gamma)$.
Thus, this relation allows us to determine the exciton contribution to the imaginary part of the conductivity, given in Eqs. (\ref{eq:cond_model_real}) and (\ref{eq:cond_model_imag}), in a more precise way.
Considering the numerical values of $\alpha(n_\gamma)$, the change from $n_\gamma\rightarrow0$ to saturation is not very large, especially in the range where we have experimental data.

\subsubsection{\label{sec:diff_model}Finite--Size Diffusion}

Because of the finite lateral size of the nanoplatelets, free charges cannot move as freely as in an infinitely extended plane.
Thus, their mobility is negatively affected due to the boundaries of the nanoplatelets.
To account for these effects, we present a diffusion model for the mobility $\mu(\omega)$ of Eqs. (\ref{eq:cond_model_real}) and (\ref{eq:cond_model_imag}).
Our approach is based on Ref. \cite{prins2006} and Chapter 9 of Ref. \cite{siebbeles2011}, and generalizes those results to two-dimensional systems with lateral sizes $L_x$ and $L_y$.
To do so we assume that free charges moving in the nanoplatelet can be described by the two-dimensional diffusion equation.
Thus, the frequency-dependent mobility can be computed from the corresponding frequency-dependent diffusion constant $D(\omega)$, which is given by

\begin{align} \label{eq:diffusion_w_2d}
    \begin{split}
        D(\omega) = D_{\text{DC}} \Bigg[ 1 &+ \sqrt{\frac{D_{\text{DC}}}{-i\omega L_x^2}} \tan \left( -\frac{1}{2} \sqrt{\frac{-i\omega L_x^2}{D_{\text{DC}}}} \right) \\
                                           &+ \sqrt{\frac{D_{\text{DC}}}{-i\omega L_y^2}} \tan \left( -\frac{1}{2} \sqrt{\frac{-i\omega L_y^2}{D_{\text{DC}}}} \right) \Bigg] ~~.\\
    \end{split}
\end{align}

\noindent This expression is found by performing the sum in Eq. (9.10) of Ref. \cite{siebbeles2011}, generalized to two dimensions.
Notice that, because of the frequency dependence in $D(\omega)$, it is not correct to assume that the total diffusion of free charges is given by setting $D_{\text{DC}} = D_{\text{DC},e} + D_{\text{DC},h}$ in Eq. (\ref{eq:diffusion_w_2d}).
Instead, it is given by $D(\omega) = D_e(\omega) + D_h(\omega)$, where each contribution is computed using a different DC value for the diffusion constant, i.e., $D_{\text{DC},e}$ and $D_{\text{DC},h}$, respectively.

Last, the frequency-dependent mobility $\mu(\omega)$ is obtained using the Einstein relation

\begin{equation} \label{eq:total_mob_2d}
    \mu(\omega) = \frac{e}{k_B T} D(\omega) = \mu_{\mathcal{R}}(\omega) + i \mu_{\mathcal{I}}(\omega)
    ~~.
\end{equation}

\noindent Since the conductivity measured in the experiments is averaged over frequencies around the peak frequency $\omega_\text{peak}$, we use $\avg{\mu(\omega)} = \mu(\omega_\text{peak})$ in the following.
The value of $\omega_\text{peak}$ is given in Table \ref{table:parameters}.

\subsection{\label{sec:mob_fit}Fitting the Electron and Hole Mobilities}

Now we are ready to gather together both the theory and the experimental data to fit the DC mobility of electrons and holes and compare them with values from the literature.
Hence, the predictions of our model are not only used to explain the measurements of interest, but can also be validated against other experiments.
In contrast to Ref. \cite{tomar2019}, we do not fit the parameters $\avg{\mu_\mathcal{R}(\omega)}$, $\avg{\mu_\mathcal{I}(\omega)}$, and $\alpha$ to the measured complex conductivity, but instead use the two supporting models from Sec. \ref{sec:supp_models} to determine them.

\begin{figure}[h]
    \begin{center}
        \includegraphics[width=\linewidth]{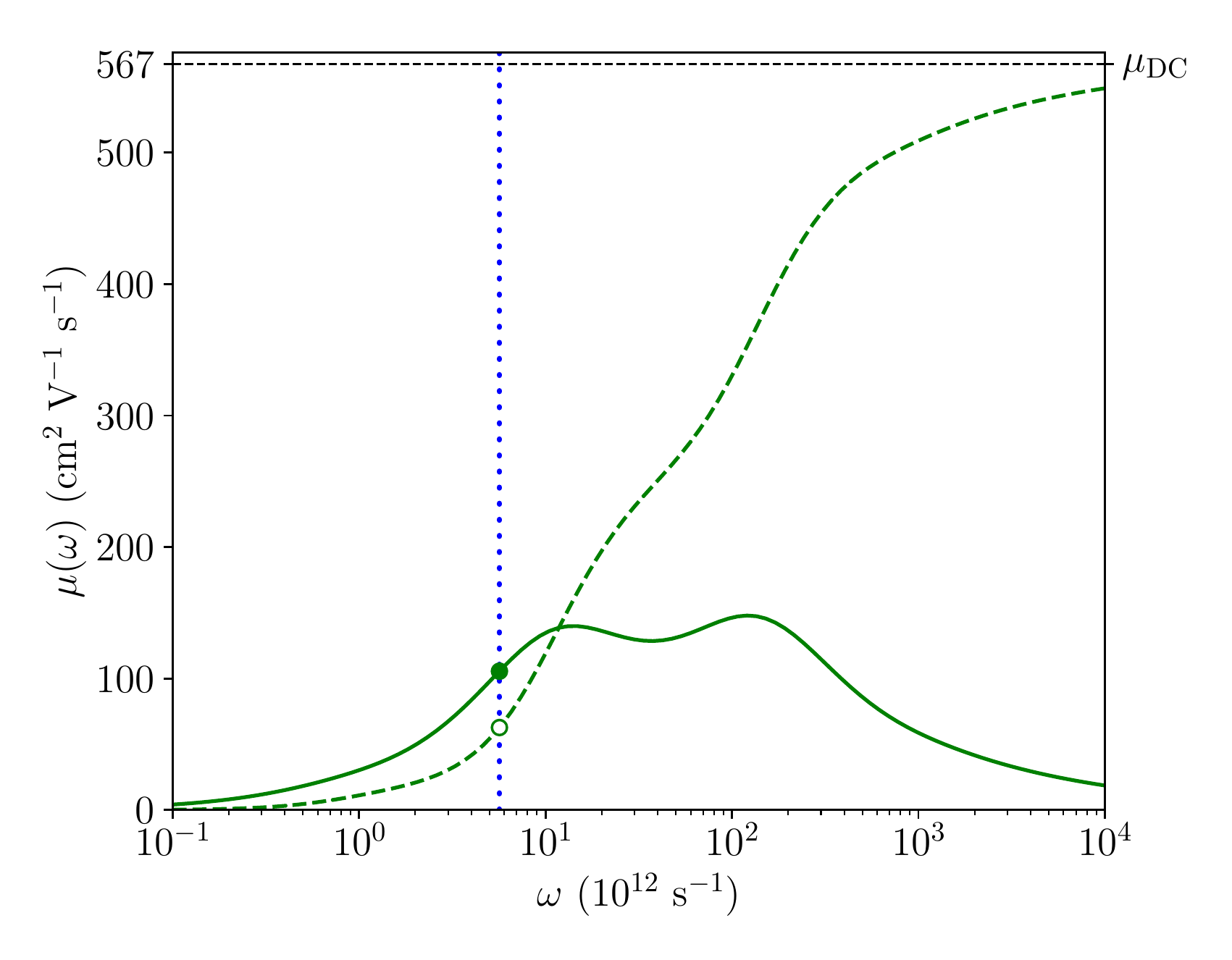}
        \caption{
            Total complex mobility of electrons and holes, that is, $\mu(\omega) \equiv \mu_{e}(\omega) + \mu_{h}(\omega)$.
            The dashed line is the real part, while the solid line is the imaginary part.
            Using $\mu_{\text{DC},e} = 527$ cm$^2$ V$^{-1}$ s$^{-1}$ and $\mu_{\text{DC},h} = 40$ cm$^2$ V$^{-1}$ s$^{-1}$, with the lateral sizes $L_x$ and $L_y$ of the system, given in Table \ref{table:parameters}.
            The vertical dotted blue line marks the peak frequency of the probe field used in the experiments, also given in Table \ref{table:parameters}.
            The green empty and full dots represent the values of $\mu_{\mathcal{R}}(\omega)$ and $\mu_{\mathcal{I}}(\omega)$, respectively, that we use in Eq. (\ref{eq:cond_model_real}) and Eq. (\ref{eq:cond_model_imag}) to obtain the curves in Fig. \ref{fig:cond_fit}.
        }
        \label{fig:mob_2d_sample}
    \end{center}
\end{figure}

Having computed the density of free charges and excitons as a function of $n_\gamma$, we can proceed to use Eqs. (\ref{eq:cond_model_real}), (\ref{eq:cond_model_imag}), and (\ref{eq:total_mob_2d}) and calculate the complex conductivity.
Because there are both electrons and holes diffusing in the nanoplatelet, we consider the contribution of electron and holes to the mobility in Eqs. (\ref{eq:cond_model_real}) and (\ref{eq:cond_model_imag}) as

\begin{equation} \label{eq:mob_cond_contr}
    \mu(\omega) = \mu_e(\omega) + \mu_h(\omega)
    ~~,
\end{equation}

\noindent where $\mu_\alpha(\omega)$ is computed using Eq. (\ref{eq:total_mob_2d}).
Figure \ref{fig:mob_2d_sample} shows the result of applying Eq. (\ref{eq:mob_cond_contr}) to our system, using the fitted DC mobilities shown in Fig. \ref{fig:cond_fit}.
Notice that the mobilities at the peak frequency $\omega_\text{peak}$, marked by a vertical dotted line, are significantly smaller than the DC values due to the finite lateral sizes of the nanoplatelets.
Each contribution introduces one DC mobility, i.e., $\mu_{\text{DC},e}$ and $\mu_{\text{DC},h}$, which we can then fit to reproduce the experimental data and compare the obtained value with literature.
Figure \ref{fig:cond_fit} shows the result of this procedure, with $\mu(\omega_\text{peak}) = 63+106i~\text{cm}^2~\text{V}^{-1}~\text{s}^{-1}$.
Comparing with the results of Ref. \cite{tomar2019}, that is, using a classical equation of state and also fitting the polarizability, they find $\mu(\omega_\text{peak}) = (54 \pm 12) + (7 \pm 5)i~\text{cm}^2~\text{V}^{-1}~\text{s}^{-1}$ and $\alpha = 3.1 \pm 0.2 \times 10^{-36}~\text{cm}^2~\text{V}^{-1}$.
Notice that because the exciton polarizability we computed is significantly lower, the imaginary part of the mobility has to be higher to compensate.
Physically, the direct fit of these three variables in Ref. \cite{tomar2019} overestimates the dephasing of the THz probe field by the polarizability of excitons, while underestimating the scattering of free charges.

Ultimately, we find in this manner the fitted DC mobilities

\begin{align}
    \mu_{\text{DC},e} &= 527~\text{cm}^2~\text{V}^{-1}~\text{s}^{-1} ~~, \\
    \mu_{\text{DC},h} &= \phantom{5}40~\text{cm}^2~\text{V}^{-1}~\text{s}^{-1}
    ~~,
\end{align}

\noindent which we can compare with Refs. \cite{madelung1999,madelung1999b} at $T=300$ K as

\begin{align}
    \mu_{\text{DC},e}^{\text{bulk}} &\simeq 600~\text{cm}^2~\text{V}^{-1}~\text{s}^{-1} ~~, \\
    \mu_{\text{DC},h}^{\text{bulk}} &\simeq \phantom{6}40~\text{cm}^2~\text{V}^{-1}~\text{s}^{-1}
    ~~.
\end{align}

\noindent Notice that in Refs. \cite{madelung1999,madelung1999b} the system is a CdSe crystal, and therefore it is representing the bulk behavior of electrons and holes.
Since our system consists of 4.5 layers, we expect the mobility to be affected by the finite layer thickness, but instead we actually find that our mobilities remain very close to the bulk ones.

\section{\label{sec:outlook}Discussion and Outlook}

In conclusion, we have introduced a complete description of the CdSe nanoplatelets involving free charges and bound exciton states in chemical (quasi)equilibrium, as introduced in Sec. \ref{sec:equilibrium}.
The system is described by the equation of state that Sec. \ref{sec:eq_state} presents, together with derivations for the exciton polarizability and mobility of free charges, given in Sec. \ref{sec:supp_models}.
Thanks to these, the conductivity model shown in Sec. \ref{sec:cond_model} only has two unknown parameters, that is, the DC mobilities of electrons and holes.
Further verification of the fitted values with the literature indicates that the insight of our model into the physics describing this system appears to be consistent.
In the following we revisit the approximations made in deriving our model, and consider the possible future directions they can lead us toward.

First, we begin by considering the differences between the Saha model, used in Ref. \cite{tomar2019}, and our model, specially with respect to Eq. (\ref{eq:id_gas_n_exc}), which is repeated below.
In addition, we have explained how the high density of excitons requires using quantum statistics, that is, the Bose-Einstein distribution $N_{BE}$.
Now we compare the density calculated using $N_{BE}$ with the result of using the Maxwell-Boltzmann distribution, as

\begin{align}
    \tag{\ref{eq:id_gas_n_exc}}
    \begin{split}
        n_{\text{exc}} = &-g_s^2 \frac{(m_e + m_h) k_B T}{2 \pi \hbar^2} \\
                         &\times \ln\left(1 - e^{\left[\mu_{\text{exc}} - E_B(n_q)\right] / k_B T}\right) ~~\text{,}
    \end{split} \\
    \label{eq:saha_exc_density}
    n_{\text{exc}}^{\text{(S)}} &= g_s^2 \frac{(m_e + m_h) k_B T}{2 \pi \hbar^2} e^{\left[\mu_{\text{exc}} - E_B(n_q)\right] / k_B T}
    ~~.
\end{align}

\noindent Hence, Eq. (\ref{eq:id_gas_n_exc}) reduces to Eq. (\ref{eq:saha_exc_density}) when the exponent $\left[\mu_{\text{exc}} - E_B(n_q)\right] / k_B T$ is very negative.
However, for a correct description of the system at a high density of excitons, that is, $n_{\text{exc}} \lambda_{\text{th}}^2 \gtrsim g_s^2$, quantum effects have to be considered.
In this regime the exciton's chemical potential is very close to its energy, and thus the mentioned approximation is no longer valid.
Notice that in Ref. \cite{tomar2019} also the screening effects are neglected by setting $E_B(n_q) = E_B(n_q=0) = E_B^\text{Cou}$.

Second, we study in more detail how excited exciton states are populated.
As we have mentioned in Sec. \ref{sec:mat_properties}, our model only treats the lowest exciton state, since contributions from the others are assumed not to be significant.
To quantify this statement we take the first excited state of the exciton, analogue to the hydrogen $2$s state, by solving Eq. (\ref{eq:eh_schr_u}).
The next most negative value of $E_B(n_q)$ that corresponds to a physical wavefunction for the exciton is the energy of this level, that we denote as $E_{B,2\text{s}}(n_q)$.
Using Eq. (\ref{eq:id_gas_n_exc}) to determine the density of this state, defined as $n_{\text{exc},2\text{s}}$, by setting $E_B(n_q)$ to $E_{B,2\text{s}}(n_q)$, results in

\begin{align}
    E_{B,2\text{s}}(n_q^\infty) &\simeq -12 ~\text{meV} ~~, \\
    \label{eq:n_exc_1_limit}
    n_{\text{exc},2\text{s}}(n_q^\infty) &\simeq 1.92 \cdot 10^{-4} ~\text{nm}^{-2} ~~.
\end{align}

\noindent However, Eq. (\ref{eq:id_gas_n_exc}) tells us that as $n_q\rightarrow n_q^\infty$, $n_{\text{exc}} \rightarrow \infty$, and any excitation will saturate to the value shown in Eq. (\ref{eq:n_exc_1_limit}).
Thus, at higher densities the amount of excited states is negligible.
At lower densities the $2$s state energy is $E_{B,2\text{s}}^\text{Cou} \simeq -21$ meV, and since $\mu_\text{exc}$ is much more negative than this the population of excited excitons is again negligible.

Finally, let us look at a different aspect of Fig. \ref{fig:cond_fit}, namely the values of $n_{\text{exc}} a_0^2$.
Since $a_0$ is the exciton Bohr radius, a density satisfying $n_{\text{exc}} a_0^2 \simeq 1$ means that there is approximately one exciton per Bohr area, and therefore at densities $n_{\text{exc}} a_0^2 \gtrsim 1$ excitons can have a substantial overlap in the nanoplatelet.
As it stands, our model neglects any possible effects coming from these overlaps.
However, we can safely assume that exciton-exciton interactions may become more relevant as the density increases and thus it is important to estimate their significance.
For this purpose, we focus on the formation of exciton complexes, analogue to molecular hydrogen H$_2$, that are known as biexcitons.
We expect that at high enough densities larger complexes, such as triexcitons, may possibly also form, but for this discussion we only consider biexcitons.
Specifically we are interested in the effect we expect them to have on the conductivity measurements that Fig. \ref{fig:cond_fit} shows.
As we have mentioned in Sec. \ref{sec:supp_models}, excitons contribute to the imaginary part of the conductivity because of their polarizability, as given in Eq. (\ref{eq:cond_model_imag}).
Thus, our goal is to estimate how the presence of biexcitons would affect the polarizability of the system.

Consider the polarizability of hydrogen molecules in three dimensions $\alpha_{\text{H}_2}$, denoted as $\alpha_{00}$ in Table VI of Ref. \cite{kolos1967}.
Compared with the polarizability of hydrogen $\alpha_{\text{H}}$ it obeys

\begin{equation}
    \label{eq:pol_h2_h_ratio}
    \alpha_{\text{H}_2} \simeq 1.21 ~ \alpha_\text{H}
    ~~.
\end{equation}

\noindent When two hydrogen atoms with polarizability $2\alpha_{\text{H}}$ bind together into a molecule, the polarizability of the system decreases to $\alpha_{\text{H}_2}$.
Therefore, the total polarizability of the system is reduced by $\alpha_{\text{H}_2}/2\alpha_{\text{H}} \simeq 60\%$.
Now, assuming that the polarizability of biexcitons behaves in a similar way to that of hydrogen molecules, binding two excitons into one biexciton decreases the total polarizability of the complex by a factor close to $1/2$.
Thus, when excitons bind into biexcitons in large quantities, the imaginary part of the conductivity decreases, since the total polarizability of the system is reduced.
Figure \ref{fig:cond_fit} does not show, within the margin of error, such behavior.
Hence it is safe to assume that, in the regime described by our model, there is not a significant population of biexcitons.

Moreover, we can estimate the density of biexcitons in the system using their measured binding energy from optical gain measurements.
In first approximation biexcitons can be described similarly to excitons, and thus their density is given by

\begin{align} \label{eq:n_x2_id_mb}
    \begin{split}
        n_{\text{exc}_2} = &- \frac{2(m_e+m_h) k_B T}{2\pi\hbar^2} \\
                           &\times \ln\left( 1 - e^{\left[2\mu_{\text{exc}} - 2 E_B(n_q) - E_{B,\text{exc}_2}(n_q)\right]/k_B T} \right) ~~,
    \end{split}
\end{align}

\noindent where $E_{B,\text{exc}_2}(n_q = 0)$ is the unscreened negative biexciton energy, that is approximately $-45$ meV \cite{tomar2019}.
To arrive at this expression we have used that the biexciton ground state is not degenerate.
Applying Eq. (\ref{eq:n_x2_id_mb}) to our system, and neglecting screening effects for simplicity, shows that at the lowest experimental densities $n_{\text{exc}_2}$ is five to ten times smaller than $n_{\text{exc}}$ and can thus be neglected, whereas at the highest experimentally achieved densities the presence of biexcitons may become noticeable.
However, changes in temperature substantially affect the density of biexcitons, for instance an increase of $200$ K reduces it by a factor of five.
Even though the conductivity measurements do not show the presence of biexcitons, we do expect that they should ultimately become more prevalent at higher densities.
Solving Eq. (\ref{eq:n_x2_id_mb}) in this high-density regime does yield that the biexciton states become significantly populated, in agreement with our expectations.

A consequence of introducing exciton-exciton interactions is that it provides a possible mechanism for breaking up excitons.
Since these interactions are only significant at high densities, above those that our model describes and that have been considered in Ref. \cite{tomar2019}, we expect that for two-dimensional systems the Mott crossover takes place in a very different regime when compared with three-dimensional systems.
For these reasons, a clear next step is to study exciton-exciton interactions in more detail, focusing on understanding the much more involved four-body problem.
In future experiments, this high-density regime may indeed be probed and our model can be updated to account for these interaction effects.
Moreover, we expect that lowering the temperature should significantly improve the exciton physics, since excitons are considerably easier to form, thus making studying their interactions much more straightforward.
Considering that in the current experiments excitons are already quantum degenerate, substantially increasing their population gives us hope for achieving superfluidity.
In conclusion, our model does not only represent an important step into a better understanding of colloidal nanoplatelets, but is hopefully also a good starting point for future theoretical and experimental efforts on exciton dynamics at high densities.

\section*{Acknowledgments}

It is a pleasure to thank Dani\"el Vanmaekelbergh, Michele Failla, and Bas Salzmann for very helpful discussions.
This work is part of the research programme TOP-ECHO with project number 715.016.002, and is also supported by the D-ITP consortium.
Both are programmes of the Netherlands Organisation for Scientific Research (NWO) that is funded by the Dutch Ministry of Education, Culture and Science (OCW).

\appendix

\section{\label{app:scr_pot_mb_2d}Screened Coulomb Potential in the Classical Limit}

To derive the screening of the Coulomb potential, we need to compute the electron-hole polarizations $\Pi_\alpha(\omega, \vec{k})$.
This derivation can be found in Secs. 8.7.1 and 8.7.2 of Ref. \cite{stoof2009}.
Performing the integral shown in Eq. (8.137) using the Maxwell-Boltzmann distribution instead of the Fermi-Dirac distribution we find

\begin{align}
    \label{eq:cl_pol_kw}
    \begin{split}
        \Pi_{\alpha}&(k, \omega; n_q) = \frac{m_\alpha}{\pi\hbar^2} \frac{1}{\sqrt{\beta\epsilon_{k}}} \\
        \times \Bigg[ &i\sqrt{\pi} \exp\left(- \beta \frac{(\hbar\omega)^2 + \epsilon_{k}^2}{4 \epsilon_{k}} \right) \sinh\left(\frac{1}{2} \beta \hbar \omega\right) \\
                      &+ F\left(\sqrt{\frac{\beta}{4\epsilon_{k}}}(-\hbar\omega+\epsilon_{k})\right) \\
                      &- F\left(\sqrt{\frac{\beta}{4\epsilon_{k}}}(-\hbar\omega-\epsilon_{k})\right) \Bigg]  e^{\beta\mu_\alpha(n_q)} \\
    \end{split}
    ~~,
\end{align}

\noindent where $\beta\equiv 1/k_B T$ and $F(x)$ is Dawson's integral (pp. 295 and 319 of Ref. \cite{abramowitz1970}), defined as

\begin{equation}
    F(x) \equiv \frac{\sqrt{\pi}}{2} e^{-x^2} \text{erf}(x)
    ~~,
\end{equation}

\noindent with $\text{erf}(x)$ the error function.
In the static limit $\omega \rightarrow 0$, Eq. (\ref{eq:cl_pol_kw}) leads to the momentum-dependent screening length

\begin{equation} \label{eq:scr_length_mb_2d}
    \lambda_{s,\alpha}^{-1}(k; n_q) = \frac{e^2}{2\epsilon_0\epsilon_r} \frac{m_\alpha}{\pi \hbar^2} \frac{2}{\sqrt{\beta\epsilon_{k}}} F\left(\frac{1}{2}\sqrt{\beta\epsilon_{k}}\right) e^{\beta\mu_\alpha(n_q)}
    ~~.
\end{equation}

\noindent Notice that we have used the property $F(-x) = -F(x)$ and introduced the correct prefactor.
In the limit $k\rightarrow\infty$ the screening length $\lambda_{s,\alpha}^{-1}(k; n_q)$ tends toward zero, and thus the screened potential reduces to the Coulomb potential.
Furthermore, taking the limit $k\rightarrow0$ results in

\begin{equation}
    \lim_{k\rightarrow0}\lambda_{s,\alpha}^{-1}(k;n_q) = \lambda_{s,\alpha}^{-1}(n_q) = \frac{e^2}{2\epsilon_0\epsilon_r} \frac{m_\alpha}{\pi\hbar^2} e^{\beta\mu_\alpha(n_q)}
    ~~,
\end{equation}

\noindent which is the same result as if we had used Eq. (\ref{eq:scr_length_2d}), with $n_\alpha(\mu)$ being the density of a classical ideal gas described by the Maxwell-Boltzmann distribution.

\section{\label{app:scr_pot_lwl_2d}Screened Coulomb Potential in the Long--Wavelength Approximation}

In Sec. \ref{sec:sat_scr} we introduced the screened potential in momentum space as

\begin{equation} \tag{\ref{eq:scr_pot_lwl_2d}}
    V_{\text{sc}}(k; n_q) = -\frac{e^2}{2 \epsilon_0 \epsilon_r} \frac{1}{k + \lambda_{s}^{-1}(n_q)}
    ~~,
\end{equation}

\noindent which we can Fourier transform back to coordinate space analytically, as

\begin{align}
    \begin{split}
        V_{\text{sc}}(r; n_q) &= \int\frac{\dif^2\vec{k}}{(2\pi)^2} ~ V_{\text{sc}}(k; n_q) e^{-i \vec{k}\cdot\vec{r}} \\
                                      &= \int_0^\infty \frac{\dif k}{2\pi}  ~ k V_{\text{sc}}(k; n_q) J_0(kr)
                                      ~~,
    \end{split}
\end{align}

\noindent where $J_0(x)$ is the Bessel function of the first kind.
Performing this integral, we find

\begin{align} \label{eq:src_pot_lwl_r_2d}
    \begin{split}
        V_{\text{sc}}&(r; n_q) = \frac{e^2}{4\pi\epsilon_0\epsilon_r} \frac{1}{\lambda_s(n_q)} \\
                     &\times \left\{ \frac{\lambda_s(n_q)}{r} + \frac{\pi}{2} \left[ Y_0\left(\frac{r}{\lambda_s(n_q)}\right) - H_0\left(\frac{r}{\lambda_s(n_q)}\right) \right] \right\}
                     ~~,
    \end{split}
\end{align}

\noindent where $H_0(x)$ is Struve's function, defined in Eq. (1) from Sec. 10.4 of Ref. \cite{watson1995}, and $Y_0(x)$ is the Bessel function of the second kind.
These special functions do not clearly show the behavior of the potential, but we can use their asymptotic expansions to derive simpler expressions.
The limits $r \gg \lambda_s(n_q)$ and $r \ll \lambda_s(n_q)$ can be found by using Eq. (2) from Sec. 10.42 of Ref. \cite{watson1995}, resulting in

\begin{widetext}
    \begin{align}
        \label{eq:scr_pot_lwl_r_limit_2d}
        V_{\text{sc}}(r; n_q) &\simeq -\frac{e^2}{4 \pi \epsilon_0 \epsilon_r}\frac{1}{\lambda_s(n_q)} \left( \frac{\lambda_s(n_q)}{r} \right)^3  , ~ r \gg \lambda_s(n_q) ~~, \\
        \label{eq:scr_pot_lwl_r_limit_long_2d}
        V_{\text{sc}}(r; n_q) &\simeq -\frac{e^2}{4 \pi \epsilon_0 \epsilon_r} \frac{1}{\lambda_s(n_q)}\left[ \frac{\lambda_s(n_q)}{r} + \ln\left(\frac{\lambda_s(n_q)}{r}\right) \right] , ~ r \ll \lambda_s(n_q) ~~.
    \end{align}
\end{widetext}

\noindent Notice that the screened potential indeed decays faster than the Coulomb potential, but only algebraically as $1/r^3$.
We can attribute this behavior to the fact that the only screening in our system comes from the two-dimensional nanoplatelet.
Therefore the fall-off is not as fast as in the Yukawa potential in three-dimensions, behaving as

\begin{equation} \label{eq:scr_pot_lwl_3d}
    V_{\text{sc}}^{\text{(3d)}}(r; n_q) = -\frac{e^2}{4 \pi \epsilon_0 \epsilon_r}\frac{1}{r} e^{-r/\lambda_s(n_q)}
    ~~,
\end{equation}

\noindent and having exponential screening.

%\nocite{*}
\bibliography{reference}% Produces the bibliography via BibTeX.

\end{document}